\documentclass[journal]{IEEEtran}
\usepackage{ucs}
\usepackage[utf8x]{inputenc}
\usepackage[cmex10]{amsmath}
\usepackage{cite, amsfonts, amssymb, amsthm, bm, bbm, graphicx, relsize, multirow, booktabs, tikz,subfigure,soul}
\usepackage[american]{babel}
\usepackage[T1]{fontenc}
\usepackage{algorithmic, algorithm}
\usepackage[multiple]{footmisc}
\usepackage{glossaries}

\newcommand{\ds}{\displaystyle}

\newtheorem{proposition}{Proposition}
\newtheorem{lemma}{Lemma}

\theoremstyle{definition}

\theoremstyle{remark}
\newtheorem{remark}{Remark}

\newcommand{\bG}{\mathbf{G}}
\newcommand{\bL}{\mathbf{L}}
\newcommand{\bA}{\mathbf{A}}
\newcommand{\bbeta}{\boldsymbol{\eta}}
\newcommand{\bx}{\boldsymbol{x}}

\makeglossaries

\newacronym{see}{SEE}{secrecy energy efficiency}
\newacronym{miso}{MISO}{multiple input single output}
\newacronym{miso-se}{MISO-SE}{multiple input single output single-antenna eavesdropper}
\newacronym{lmmse}{LMMSE}{linear minimum mean square error}
\newacronym{d2d}{D2D}{device-to-device}
\newacronym{p2p}{P2P}{point-to-point}
\newacronym{mac}{MAC}{multiple-access channel}
\newacronym{bc}{BC}{broadcast channel}
\newacronym{ic}{IC}{interference channel}
\newacronym{imac}{IMAC}{interference multiple access channel}
\newacronym{ibc}{IBC}{interference broadcast channel}
\newacronym{mimo}{MIMO}{multiple-input multiple-output}
\newacronym{mimo-me}{MIMO-ME}{multiple input multiple output multiple-antenna eavesdropper}
\newacronym{siso}{SISO}{single-input single-output}
\newacronym{sc}{SC}{single-carrier}
\newacronym{mc}{MC}{multi-carrier}
\newacronym{ofdma}{OFDMA}{orthogonal frequency division multiple access}
\newacronym{af}{AF}{amplify-and-forward}
\newacronym{df}{DF}{decode-and-forward}
\newacronym{cf}{CF}{compress-and-forward}
\newacronym{mwrc}{MWRC}{multi-way relay channel}
\newacronym{pde}{PDE}{partial data exchange}
\newacronym{fde}{FDE}{full data exchange}
\newacronym{iid}{i.i.d.\@}{independent and identically distributed}
\newacronym{awgn}{AWGN}{additive white Gaussian noise}
\newacronym{awg}{AWG}{additive white Gaussian}
\newacronym{sic}{SIC}{successive interference cancellation}
\newacronym{dpc}{DPC}{dirty paper coding}
\newacronym{snr}{SNR}{signal-to-noise ratio}
\newacronym{sinr}{SINR}{signal to interference plus noise ratio}
\newacronym{ber}{BER}{bit error rate}
\newacronym{zf}{ZF}{zero-forcing}
\newacronym{mmse}{MMSE}{minimum mean square error}
\newacronym{sud}{SUD}{single user decoding}
\newacronym{dof}{DoF}{degrees of freedom}
\newacronym{gdof}{GDoF}{generalized degrees of freedom}
\newacronym{nnc}{NNC}{noisy network coding}
\newacronym{dmn}{DMN}{discrete memoryless network}
\newacronym{csi}{CSI}{channel state information}
\newacronym{ee}{EE}{energy efficiency}
\newacronym{ian}{IAN}{treating interference as noise}
\newacronym{snd}{SND}{simultaneous non-unique decoding}
\newacronym{brd}{BRD}{best response dynamics}
\newacronym{br}{BR}{best response}
\newacronym{ne}{NE}{Nash equilibrium}
\newacronym{lhs}{LHS}{left-hand side}
\newacronym{rhs}{RHS}{right-hand side}
\newacronym{gee}{GEE}{global energy efficiency}
\newacronym{wsee}{WSEE}{weighted sum energy efficiency}
\newacronym{wpee}{WPEE}{weighted product energy efficiency}
\newacronym{wmee}{WMEE}{weighted minimum energy efficiency}
\newacronym{kkt}{KKT}{Karush Kuhn Tucker}
\newacronym{pc}{PC}{pseudo-concave}
\newacronym{qc}{QC}{quasi-concave}
\newacronym{ql}{QL}{quasi-linear}
\newacronym{pl}{PL}{pseudo-linear}
\newacronym{spc}{SPC}{strictly pseudo-concave}
\newacronym{sqc}{SQC}{strictly quasi-concave}
\newacronym{lfp}{LFP}{linear fractional problem}
\newacronym{clfp}{CLFP}{concave-linear fractional problem}
\newacronym{ccfp}{CCFP}{concave-convex fractional problem}
\newacronym{mmfp}{MMFP}{max-min fractional problem}
\newacronym{sorp}{SoRP}{sum-of-ratios problem}
\newacronym{porp}{PoRP}{product-of-ratios problem}
\newacronym{qos}{QoS}{quality of service}
\newacronym{evd}{EVD}{eigenvalue decomposition}
\newacronym{svd}{SVD}{singular value decomposition}
\newacronym{skee}{SKEE}{Secret-key energy efficiency}
\newacronym{an}{AN}{artificial noise}

\begin{document}

\bstctlcite{IEEEexample:BSTcontrol}

\title{User-Centric 5G Cellular Networks: Resource Allocation and Comparison with the Cell-Free Massive MIMO Approach}
\author{Stefano Buzzi, {\em Senior Member}, {\em IEEE}, Carmen D'Andrea, and Alessio Zappone {\em Senior Member, {\em IEEE}}
\thanks{This paper was partly presented at the 21th International ITG Workshop on Smart Antennas, Berlin (Germany), March 2017, and at the 28th Annual IEEE International Symposium on Personal, Indoor and Mobile Radio Communications, Montreal (Canada), October 2017. This research has been supported by the Italian Ministry of Education and Research, under the program "Dipartimenti di Eccellenza 2018-2022".}
\thanks{S. Buzzi and C. D'Andrea are with the Department of Electrical and Information Engineering, University of Cassino and Lazio Meridionale, I-03043 Cassino, Italy (buzzi@unicas.it, carmen.dandrea@unicas.it).}
\thanks{Alessio Zappone is with the LANEAS group of the L2S, CentraleSupelec, CNRS, UnivParisSud, Universit\'e Paris-Saclay, 91192 Gif-sur-Yvette, France. (alessio.zappone@l2s.centralesupelec.fr).}}
\maketitle
\thispagestyle{empty}
\begin{abstract}
Recently, the so-called cell-free  (CF) Massive MIMO architecture has been introduced, wherein a very large number of distributed access points (APs) simultaneously and jointly serve a much smaller number of mobile stations (MSs). The paper extends the CF approach to the case in which both the APs and the MSs are equipped with multiple antennas, proposing a beamfoming scheme that, relying on the channel hardening effect,  does not require channel estimation at the MSs. We contrast the CF massive MIMO approach with a user-centric (UC) approach wherein each MS is served only by a limited number of APs. Since far APs experience a bad SINR, it turns out that they are quite unhelpful in serving far users, and so, the UC approach, while requiring less backhaul overhead with respect to the CF approach, is shown here to achieve better performance results, in terms of achievable rate-per-user, for the vast majority of the MSs in the network. Furthermore, in the paper we propose two power allocation strategy for the uplink and downlink, one aimed at maximizing the overall data-rate and another aimed at maximizing system fairness.
\end{abstract}


\section{Introduction}

Massive multiple-input multiple-output (MIMO), introduced by Marzetta in his pioneering paper \cite{Marzetta10} is a promising 5G wireless access technology that can provide high throughput with simple signal processing \cite{whatwillbe}. Massive antenna at the base stations can be deployed in co-located or distributed setups. In co-located Massive MIMO all the antennas are located in a compact area and this architecture has the advantage of low backhaul requirements. In distributed Massive MIMO systems, instead, the antennas are spread out over a large area; this architecture has the advantage of efficiently exploiting mascroscopic  diversity against the shadow fading, so these systems can potentially offer much higher probability of coverage than collocated Massive MIMO \cite{ZhouDAS2003}, at the cost of increased backhaul requirements. Additionally, the distributed layout permits alleviating the cell-edge problem, since it considerably lowers the probability that a user happens to be situated far from every system access-point, permitting thus to achieve a better fairness and service uniformity across mobile users.
In \cite{DAS_Heath2013}, the viability of using distributed antennas in multi-cell systems for massive MIMO on the uplink is investigated for a particular spatial correlation channel model. In \cite{FengVirtualMIMO2013} the authors focus on the downlink of a multicell distributed antenna system assuming that only the slowly-varying large-scale channel state is required at the transmitter and they explore the performance gain that can be achieved by coordinated transmissions for a virtual MIMO system. 
One of the drawbacks of such virtual MIMO systems is the heavy backhaul requirements, since, besides data symbols, also the channel estimates and the beamforming schemes are to be shared with the central processing unit (CPU). 

Recently, a CF massive MIMO architecture has been introduced \cite{ngo2015cell,Ngo_CellFree2017}  where a very large number of distributed multiple-antenna access points (APs) serve many single-antenna mobile stations (MSs) in the same time-frequency resource. All APs re connected to a CPU and cooperate via a backhaul network, serving all the users via time-division duplex (TDD) operation. In a CF massive MIMO system there are actually no cells or cell boundaries, and the system is such that 
the backhaul is used to transmit data-symbols on the downlink and sufficient statistics to the CPU on the uplink, to enable uplink data detection; otherwise stated, channel estimates at the APs are not forwarded to the CPU and the beamformers are computed locally. 
The authors of\cite{Ngo_CellFree2017} show that the CF approach provides better performance than a small-cell system in terms of 95\%-likely per-user throughput, thus confirming that the scheme is effective in alleviating the cell-edge user problem and in providing a more uniform service across users. CF massive MIMO is a recent research topic that however has been gaining a strong momentum in the last few years. The paper \cite{cell-free_downlinkpilots} has shown that some performance improvement can be obtained in low density networks by using downlink pilots, while the paper \cite{cell-free-precoding}, instead, analyzes the performance improvements granted by the use of a zero-forcing precoder in the downlink: although the gains are from five to ten-fold, the zero-forcing precoder requires centralized computations at the CPU and increased backhaul overhead. 
Zero-forcing precoding is again considered in \cite{Nguyen_CL}, wherein it is coupled with a power control algorithm aimed at  maximizing the energy efficiency of CF massive MIMO considering the backhaul power consumption and the imperfect channel state information. 
In \cite{nayebi2016performance}, the uplink performance of CF systems
is investigated using minimum mean square error (MMSE) processing for the case in which 
the energy efficiency of CF massive MIMO is to be maximized considering the backhaul power consumption and the imperfect channel state information.
The energy-efficiency of CF massive MIMO systems is also considered in \cite{Ngo_EE_cellfree}, which proposes a power allocation algorithm aiming at maximizing the
total energy efficiency, subject to a per-user spectral efficiency
constraint and a per-AP power constraint. The power allocation strategies here are simplified by the fact that single-antenna transceivers are considered both at the APs and at the MSs, which permit skipping $\log \det[\cdot]$ functions in the achievable rate formulas. 
In order to reduce the backhaul overhead, the paper \cite{huang2017compute} considers instead a coded CF massive MIMO system, and investigates the performance of a compute $\&$ forward mechanism for the uplink, wherein each AP attempts to use an integer linear combination of the
codewords to represent the scaled received signal to be sent to the CPU. 
The finite backhaul capacity is also considered in \cite{bashar2018cell}, which studies the case in which quantized version of the
estimated channel and the quantized received signal are available
at the CPU, and the case when only the
quantized version of the combined signal with maximum ratio
combining  detector is available at the CPU.

It should be noted that all the cited papers consider the case in which both the APs and the MSs are equipped with a single-antenna, the only exception being reference \cite{bashar2018cell}, which considers multiple antennas at the APs. The extension of the CF massive MIMO architecture to the case in which also the MSs are equipped with multiple antennas is not trivial since no channel estimation is performed at the MSs, and so no channel-dependent beamforming scheme can be used there. 

One critical point of CF massive MIMO systems is the fact that all the APs serve all the MSs in the system. This assumption may lead to some inefficiencies in the system as the size of the considered area grows: indeed, it appears clearly pointless to wast power and computational resources at an AP to decode MSs that are very far and that are presumably received with a very low Signal-to-Interference plus Noise Ratio (SINR). To overcome this limitation, the 
a user-centric (UC) distributed massive MIMO system has been introduced, still for single-antenna APs and MSs, in \cite{buzzi_CFUC2017}; in the UC approach, each MS is served not by all the APs in the system, but just by the ones that are in the neighborhood. 
The UC approach, while being much simpler than the CF one and less hungry of backhaul bandwidth, was shown in \cite{buzzi_CFUC2017} to provide a larger achievable rate-per-user to the majority of the MSs in the system.

Following on such a track, and building upon the conference papers \cite{buzzi2017user, Buzzi_Zappone_PIMRC2017}, this paper provides a thorough comparison of the UC and CF approaches, considering the  case in which the MSs and the APs are equipped with multiple antennas. We propose a beamforming scheme that does not require channel estimation at the MSs; rather the proposed scheme exploits the channel hardening effect due to the large number of antennas in order to perform coherent data reception at the MSs. We propose a simple and low complexity pilot matched channel estimation strategy implemented at each APs and we consider maximum-length-sequences (pseudo-noise) as pilots at each MS in the uplink training phase. Channel inversion beamforming is proposed here as a generalization of the conjugate beamforming applied in the single-antenna case, and, again, no channel estimates and beamforming matrices are propagated through the backhaul network.
Furthermore, we propose two power allocation strategies for the uplink and the downlink, both for the CF and the UC case. The first one is a sum-rate maximizing power allocation strategy, aimed at maximizing performance of the system in terms of overall data-rate and the second one is a minimum-rate maximizing power allocation, aimed at maximizing performance of the system in terms of fairness. We contrast the results obtained with the power allocation strategies proposed in the paper with the case of uniform power allocation, i.e. all the APs and the MSs transmit data with the maximum power available, and we consider the CF and UC approaches with pilot matched channel estimation and with perfect channel state information (CSI). Results will show that the UC approach generally outperforms the CF one, especially on the uplink.

The remainder of this paper is organized as follows. Next Section contains the description of the considered system model. Section \ref{Comm_protocol_section} is devoted to the illustration of the comunication protocol, composed by uplink training, downlink data transmission and uplink data transmission, for both CF and UC approaches. In Section \ref{Downlink_power_control_section} we report the performance measures and the two power control strategies proposed for the downlink. Section \ref{Uplink_power_control_section} contains the two power control strategies proposed for the uplink and Section \ref{Numerical_section} contains the numerical results. Finally, concluding remarks are given in Section \ref{conclusions_section}.

\textit{Notation:}In this paper we use the following notation: $\mathbf{A}$ is a matrix; $\mathbf{a}$ is a vector; $a$ is a scalar. The operators $(\cdot)^T$, $(\cdot)^{-1}$, and $(\cdot)^H$ stand for transpose, inverse and, conjugate transpose, respectively. The determinant of the matrix $\mathbf{A}$ is denoted as $|\mathbf{A}|$ and $\mathbf{I}_P$ is the $P \times P$ identity matrix.

\section{System model} \label{System_model_section}
We consider an area with $K$ MSs and $M$ APs. MSs and APs are randomly located. 
The $M$ APs are connected by means of a backhaul network to a central processing unit (CPU) wherein data-decoding is performed. 
In keeping with the approach of \cite{ngo2015cell,Ngo_CellFree2017}, all communications take place on the same frequency band; uplink and downlink are separated through time-division-duplex (TDD); the coherence interval is thus divided into three phases: (a) uplink channel estimation, (b) downlink data transmission, and (c) uplink data transmission. In phase (a) the MSs send pilot data in order to enable channel estimation at the APs. 
In phase (b) APs use channel estimates to perform channel-matched beamforming and send data symbols on the downlink; while in the CF architecture APs send data to all the MSs in the system, in the UC approach APs send data only to a subset of the MSs in the system. 
Finally, in phase (c) MSs send uplink data symbols to the APs; while in the CF architecture all the APs participate to the decoding of the data transmitted by all the MSs, in the UC approach APs just decode the data from the nearby MSs. The procedure for the selection of the MSs to serve will be specified in the following section.
No pilots are transmitted on the downlink and  no channel estimation is performed at the MSs: data decoding takes place on the downlink relying on the fact that in TDD the downlink channel is the reciprocal of the uplink channel
\footnote{According to \cite{Ngo_CellFree2017}, the channel reciprocity is also ensured by perfect hardware chain calibration, whose feasibility has been recently shown in \cite{kaltenberger2010relative}.}
and on the channel hardening effect due to many transmitting APs. 
In the following, we denote by $N_{\rm MS}$ and by $N_{\rm AP}$ the number of antennas at the MSs and at the APs, respectively. 

\subsection{Channel model}
We denote by the $(N_{\rm AP} \times N_{\rm MS})$-dimensional matrix $\mathbf{G}_{k,m}$ the channel between the $k$-th MS and the $m$-th AP. We have
\begin{equation}
\mathbf{G}_{k,m}=\beta_{k,m}^{1/2} \mathbf{H}_{k,m} \; ,
\label{channel_model}
\end{equation}
with $\beta_{k,m}$ a scalar coefficient modeling the channel shadowing effects and 
$\mathbf{H}_{k,m}$ an $(N_{\rm AP} \times N_{\rm MS})$-dimensional matrix whose entries are i.i.d ${\cal CN}(0,1)$ RVs.
For the path loss and the shadow fading correlation models we use the ones reported in \cite{Ngo_CellFree2017}. 
The large scale coefficient $\beta_{k,m}$ in \eqref{channel_model} models the path loss and shadow fading, according to 
\begin{equation}
\beta_{k,m}= 10^{\frac{\text{PL}_{k,m}}{10}} 10^{\frac{\sigma_{\rm sh}z_{k,m}}{10}},
\end{equation}
where $\text{PL}_{k,m}$ represents the path loss (expressed in dB) from the $k$-th MS to the $m$-th AP, and $10^{\frac{\sigma_{\rm sh}z_{k,m}}{10}}$ represents the shadow fading with standard deviation $\sigma_{\rm sh}$, while $z_{k,m}$ will be specified later.
For the path loss we use the following three slope path loss model \cite{tang2001mobile}:
\begin{equation}
\text{PL}_{k,m}=\left\lbrace 
\begin{array}{lll}
-L-35 \log_{10}\left(d_{k,m}\right), & & \text{if} \; d_{k,m}>d_1 \\
-L - 10 \log_{10}\left(d_1^{1.5} d_{k,m}^{2}\right), & & \text{if} \; d_0< d_{k,m}\leq d_1 \\
-L - 10 \log_{10}\left(d_1^{1.5} d_{0}^{2}\right), & & \text{if} \; d_{k,m}<d_0
\end{array} \right. ,
\label{path_loss}
\end{equation} 
where $d_{k,m}$ denotes the distance between the $m$-th AP and the $k$-th user, $L$ is
\begin{equation}
\begin{array}{lll}
L=& 46.3+33.9\log_{10}\left(f\right)-13.82\log_{10}\left(h_{\rm AP}\right)- 
\\ & \left[1.11\log_{10}\left(f\right)-0.7\right]h_{\rm MS} +1.56\log_{10}\left(f\right)-0.8,
\end{array}
\end{equation}
$f$ is the carrier frequency in MHz, $h_{\rm AP}$ and $h_{\rm MS}$ denotes the AP and MS antenna heights, respectively.
In real-world scenarios, transmitters and receivers that are in close vicinity of each other may be surrounded by common obstacles, and hence, the shadow fading RVs are correlated; for the shadow fading coefficient we thus use a model with two components \cite{wang2008joint}
\begin{equation}
z_{k,m}=\sqrt{\delta}a_m+\sqrt{1-\delta}b_k, \; \; m=1, \ldots, M, \; k=1,\ldots,K,
\label{shadowing}
\end{equation}
where $a_m \sim \mathcal{N}(0,1)$ and $b_k \sim \mathcal{N}(0,1)$ are independent RVs, and $\delta, \; 0\leq \delta \leq 1$ is a parameter.
The covariance functions of $a_m$ and $b_k$ are given by:
\begin{equation}
E\left[a_m a_{m'}\right]=2^{-\frac{d_{\rm{AP}(m,m')}}{d_{\rm decorr}}} \; \; \; \; 
E\left[b_k b_{k'}\right]=2^{-\frac{d_{\rm{MS}(k,k')}}{d_{\rm decorr}}},
\label{shadowing_corr}
\end{equation}
where $d_{\rm{AP}(m,m')}$ is the geographical distance between the $m$-th and $m'$-th APs, $d_{\rm{MS}(k,k')}$ is the geographical distance between the $k$-th and the $k'$-th MSs. The parameter $d_{\rm decorr}$  is a decorrelation distance which depends on the environment, typically this value is in the range 20-200 m.

\section{The communication protocol for the CF and UC approaches} \label{Comm_protocol_section}
As already discussed, the communication procedure is made of three different phases, (a) uplink training, (b) downlink data transmission, and (c) uplink data transmission. The overall duration of these three phases must not exceed the channel coherence time, thus implying that these three phases must be sequentially repeated with a frequency larger than the channel Doppler spread. 

\subsection{Uplink training}
During this phase the MSs send uplink training pilots in order to permit channel estimation at the APs.  This phase is the same for both the UC and CF approaches.
We denote by $\tau_c$ the length (in samples) of the channel coherence time, and by $\tau_p$ the length (in samples) of the uplink training phase. Of course we must ensure that $\tau_p < \tau_c$. 
Denote by $\mathbf{\Phi}_k \in {\cal C}^{N_{\rm MS}\times \tau_p}$ the pilot sequence sent by the $k$-th MS, and assume that $\|\mathbf{\Phi}_k\|^2_F=1$. 
The signal received at the $m$-th AP in the $n$-th signaling time is represented by the following $N_{\rm AP}$-dimensional vector:
\begin{equation}
\mathbf{y}_m(n)= \ds \sum_{k=1}^K \ds \sqrt{p_k} \mathbf{G}_{k,m}\mathbf{\Phi}_k(:,n) + \mathbf{w}_m(n) \; ,
\end{equation}
with $\sqrt{p}_k$ the user $k$ transmit power during the training phase.
Collecting all the observable vectors  $\mathbf{y}_m(n)$, for $n=1, \ldots, \tau_p$ into the $(N_{\rm AP} \times \tau_p)$-dimensional matrix $\mathbf{Y}_m$, it is easy to show that:
\begin{equation}
\mathbf{Y}_m = \ds \sum_{k=1}^K \ds \sqrt{p_k} \mathbf{G}_{k,m}\mathbf{\Phi}_k + \mathbf{W}_m \; .
\end{equation}
In the above equation the matrix $\mathbf{W}_m$ is $(N_{\rm AP} \times \tau_p)$-dimensional and contains the thermal noise contribution and out-of-cell interference at the $m$-th AP; its entries are assumed to be i.i.d. ${\cal CN}(0, \sigma^2_w)$ RVs.
Based on the observable matrix $\mathbf{Y}_m$, the $m$-th AP performs estimation of the channel matrices 
$\left\{\mathbf{G}_{k,m}\right\}_{k=1}^K$. We assume here simple pilot-matched (PM) single-user channel estimation for the sake of simplicity (more sophisticated channel estimation schemes might however be considered) and we assume knowledge of MSs transmit powers $\left\{p_k\right\}_{k=1}^K$. The estimate,  
$\widehat{\mathbf{G}}_{k,m}$ say, of the channel matrix ${\mathbf{G}}_{k,m}$ is obtained as
\begin{equation}
\begin{array}{lll}
\widehat{\mathbf{G}}_{k,m}= & \ds \frac{1}{\sqrt{p_k}}\mathbf{Y}_m \mathbf{\Phi}_k^H=\ds 
{\mathbf{G}}_{k,m} \mathbf{\Phi}_k \mathbf{\Phi}_k^H + \\ &
\ds \sum_{j=1, j \neq k}^K \ds \sqrt{\frac{p_j}{p_k}} \mathbf{G}_{j,m} \mathbf{\Phi}_j \mathbf{\Phi}_k^H +\frac{1}{\sqrt{p_k}}\mathbf{W}_m \mathbf{\Phi}_k^H \; .
\label{eq:downlink_channel_estimate}
\end{array}
\end{equation}
Estimation \eqref{eq:downlink_channel_estimate} must be made in all the APs (i.e., for all the values of $m=1, \ldots, M$) for all the values of $k=1, \ldots, K$. If the rows of the matrices $\mathbf{\Phi}_1, \ldots, \mathbf{\Phi}_K$ are pairwisely orthogonal (i.e. $\mathbf{\Phi}_k \mathbf{\Phi}_j= \mathbf{I}_{N_{\rm MS}} \delta_{i,k}$, for all $i, k$), then Eq. \eqref{eq:downlink_channel_estimate} simplifies to 
\begin{equation}
\widehat{\mathbf{G}}_{k,m}=\ds \frac{1}{\sqrt{p_k}}\mathbf{Y}_m \mathbf{\Phi}_k^H=
{\mathbf{G}}_{k,m} +\ds \frac{1}{\sqrt{p_k}} \mathbf{W}_m \mathbf{\Phi}_k^H \; ,
\label{eq:downlink_channel_estimate_simplified}
\end{equation}
and thermal noise is the only disturbance  impairing the channel estimate. 
A necessary condition for this to happen is however $\tau_p \geq KN_{\rm MS}$, a relation that usually is not verified in practical scenarios due to the fact that $\tau_p$ must be a fraction of the channel coherence length. As a consequence, almost orthogonal pilot sequences are usually employed. In this paper, we assume that the pilot sequences assigned to each user are mutually orthogonal, so that $\mathbf{\Phi}_k \mathbf{\Phi}_k^H=\mathbf{I}_{N_{\rm MS}}$, while, instead, pilot sequences from different users are non-orthogonal. As a consequence, Eq. \eqref{eq:downlink_channel_estimate} is actually expressed as:
\begin{equation}
\widehat{\mathbf{G}}_{k,m}=
{\mathbf{G}}_{k,m}  +
\sum_{j=1, j \neq k}^K \ds \sqrt{\frac{p_j}{p_k}} \mathbf{G}_{j,m} \mathbf{\Phi}_j \mathbf{\Phi}_k^H + \ds \frac{1}{\sqrt{p_k}} \mathbf{W}_m \mathbf{\Phi}_k^H \; ,
\label{eq:downlink_channel_estimate2}
\end{equation}
which clearly shows that the channel estimate is degraded not only by noise, but also by the pilots from the other users, an effect which is well-known to be named pilot contamination.

\subsection{Downlink data transmission}
After that each AP has obtained estimates of the channel matrix from all the MSs in the system, the downlink data transmission phase begins. 
The APs treat the channel estimates as the true channels, and channel inversion  beamforming is performed to transmit data to the MSs. 
The objective of this beamforming scheme is to ensure that the MSs will be able to receive data with no information on the channel state. 
Denoting by $P_k$ the multiplexing order (i.e., the number of simultaneous data-streams) for user $k$,  and by $\mathbf{x}_k^{\rm DL}(n)$ the $P_k$-dimensional unit-norm vector containing the $k$-th user data symbols to be sent in the $n$-th sample time, and letting $\mathbf{L}_k=\mathbf{I}_{P_k} \otimes \mathbf{1}_{N_{\rm MS}/P_k}$, the downlink precoder at the $m$-th AP for the $k$-th MS is expressed as 
\begin{equation}
\mathbf{Q}_{k,m}=\widehat{\mathbf{G}}_{k,m} \left(\widehat{\mathbf{G}}_{k,m}^H \widehat{\mathbf{G}}_{k,m} \right)^{-1} \mathbf{L}_k \; .
\label{eq:downlink_precoder}
\end{equation}
\subsubsection{CF massive MIMO architecture} 
In the CF architecture all the APs communicate with all the MSs in the systems, so the signal transmitted by the $m$-th AP in the $n$-th interval is the following $N_{\rm AP}$-dimensional vector
\begin{equation}
\mathbf{s}_m^{\rm cf}(n)=\ds \sum_{k=1}^K \ds \sqrt{\eta_{k,m}^{\rm DL,cf}} \mathbf{Q}_{k,m} \mathbf{x}_k^{\rm DL}(n) \; ,
\end{equation}
with $\eta_{k,m}^{\rm DL,cf}$ a scalar coefficient ruling the power transmitted by the $m$-th AP for the $k$-th MS. 
The generic $k$-th MS receives signal contributions from all the APs; the observable vector is expressed as
\begin{equation}
\begin{array}{llll}
\mathbf{r}_k^{\rm cf}(n)& =& \ds \sum_{m=1}^M
\mathbf{G}_{k,m}^H \mathbf{s}_m^{\rm cf}(n) + \mathbf{z}_k(n) \\
& =& \ds \sum_{m=1}^M \ds \sqrt{\eta_{k,m}^{\rm DL,cf}} \mathbf{G}_{k,m}^H  \mathbf{Q}_{k,m} \mathbf{x}_k^{\rm DL}(n) + \\ & & 
\ds \sum_{m=1}^M \sum_{j=1, j \neq k}^K  \ds \sqrt{\eta_{j,m}^{\rm DL,cf}}\mathbf{G}_{k,m}^H   \mathbf{Q}_{j,m} \mathbf{x}_j^{\rm DL}(n) + \mathbf{z}_k(n) \; .
\end{array}
\label{eq:received_data_MS}
\end{equation}
In \eqref{eq:received_data_MS}, the $N_{\rm MS}$-dimensional vector $\mathbf{z}_k(n)$,
modelled as i.i.d. ${\cal CN}(0,\sigma^2_z)$ RVs,  represents the thermal noise and out-of-cluster interference at the $k$-th MS. Based on the observation of the vector $\mathbf{r}_k^{\rm cf}(n)$, a soft estimate of the data symbols 
$\mathbf{x}_k^{\rm DL}(n)$ is obtained at the $k$-th MS as
\begin{equation}
\widehat{\mathbf{x}}_k^{\rm DL,cf}(n)= \mathbf{L}_k^H \mathbf{r}_k^{\rm cf}(n) \; .
\label{Est_DL_cf1}
\end{equation}
Note that no channel estimation is performed at the MSs; the beamformers $\mathbf{L}_k$ have a fixed structure 
independent of the channel realization, so that  the entries of the observation vector are partitioned in  $P_k$ groups and a coherent sum is made within each group.

\subsubsection{UC massive MIMO architecture}
In the user-centric approach, we assume that the APs communicate only with the closest MSs. In order to define a measure for the closeness of the MSs, several procedures can be conceived. 
One possible strategy is that each AP computes the average Frobenius norm of the estimated channels  for all the MSs, i.e.:
\begin{equation}
\mathbf{\bar{G}}_{m}=\frac{1}{K}\sum_{k=1}^K{\|\widehat{\mathbf{G}}_{k,m}\|_F},
\end{equation}
and will serve only the APs whose channel has a Frobenius norm larger than the computed average value.
Another possible approach is that each AP sorts these estimates in descending Frobenius norm order and serves only the $N$ MSs with the strongest channel, with $N$ a proper design parameter. In this paper we will present numerical results using this latter strategy. 
We denote by $\mathcal{K}(m)$ the set of MSs served by the $m$-th AP. 
Given the sets $\mathcal{K}(m)$, for all $m=1, \ldots, M$,  we can define the set $\mathcal{M}(k)$  of the APs that communicate with the $k$-th user:
\begin{equation}
\mathcal{M}(k)=\{ m: \,  k \in \mathcal{K}(m) \}
\end{equation}
So, in this case, the signal transmitted by the $m$-th AP in the $n$-th interval is the following $N_{\rm AP}$-dimensional vector
\begin{equation}
\mathbf{s}_m^{\rm uc}(n)=\ds \sum_{k\in{\cal K}(m)} \ds \sqrt{\eta_{k,m}^{\rm DL,uc}} \mathbf{Q}_{k,m} \mathbf{x}_k^{\rm DL}(n) \; ,
\end{equation}
with $\eta_{k,m}^{\rm DL,uc}$, again, a scalar coefficient ruling the power transmitted by the $m$-th AP. 
The generic $k$-th MS receives signal contributions from all the APs; the observable vector is expressed as
\begin{equation}
\begin{array}{llll}
\mathbf{r}_k^{\rm uc}(n)& = & \ds \sum_{m=1}^M
\mathbf{G}_{k,m}^H \mathbf{s}_m^{\rm uc}(n) + \mathbf{z}_k(n) \\
& = &\ds \sum_{m\in{\cal M}(k)} \ds \sqrt{\eta_{k,m}^{\rm DL,uc}} \mathbf{G}_{k,m}^H  \mathbf{Q}_{k,m} \mathbf{x}_k^{\rm DL}(n) + \\ & & 
\ds \sum_{j=1, j \neq k}^K \sum_{m\in{\cal M}(j)}   \ds \sqrt{\eta_{j,m}^{\rm DL,uc}}\mathbf{G}_{k,m}^H   \mathbf{Q}_{j,m} \mathbf{x}_j^{\rm DL}(n) + \mathbf{z}_k(n)\; .
\end{array}
\label{eq:received_data_MS_UC}
\end{equation}
In \eqref{eq:received_data_MS_UC}, the $N_{\rm MS}$-dimensional vector $\mathbf{z}_k(n)$   represents the thermal noise and out-of-cluster interference at the $k$-th MS, and is modeled
 as i.i.d. ${\cal CN}(0,\sigma^2_z)$ RVs. 
 Based on the observation of the vector $\mathbf{r}_k^{\rm uc}(n)$, a soft estimate of the data symbols 
$\mathbf{x}_k^{\rm DL}(n)$ is obtained at the $k$-th MS as
\begin{equation}
\widehat{\mathbf{x}}_k^{\rm DL,uc}(n)= \mathbf{L}_k^H \mathbf{r}_k^{\rm uc}(n) \; .
\label{Est_DL_uc1}
\end{equation}

\subsection{Uplink data transmission}
The final phase of the communication protocol consists of the uplink data transmission.
Since the MSs do not perform channel estimation, they just send their data symbols using the already defined trivial beamformer $\mathbf{L}_k$.  Basically, this corresponds to partition the MS antennas in as many disjoint subsets as the multiplexing order, and to use all the antennas in each same subset to transmit the same data symbol. 
We denote by $\mathbf{x}^{\rm UL}_k(n)$ the $P_k$-dimensional data vector to be transmitted by the $k$-th user in the $n$-th sample time. The signal received at the $m$-th AP in the $n$-th time sample is  an $N_{\rm AP}$-dimensional vector expressed as 
\begin{equation}
\mathbf{\bar{y}}_m(n)=\ds \sum_{k=1}^K \ds \sqrt{\eta_{k}^{\rm UL}} \mathbf{G}_{k,m} \mathbf{L}_k \mathbf{x}^{\rm UL}_k(n) +
\mathbf{w}_m(n) \; ,
\end{equation}
with $\eta_{k}^{\rm UL}$ is the uplink transmit power of the $k$-th MS.

\subsubsection{CF massive MIMO architecture} 
In the case of CF MIMO, all the APs participate to the decoding of the data sent by all the MSs. 
The $m$-th AP, thus, forms, for each $k=1, \ldots, K$,  the following statistics
\begin{equation}
\begin{array}{lll}
\widetilde{\mathbf{y}}_{m,k}(n)& =  \left(\mathbf{L}_k^H \widehat{\mathbf{G}}_{k,m}^H \widehat{\mathbf{G}}_{k,m}
\mathbf{L}_k\right)^{-1}   \mathbf{L}_k^H \widehat{\mathbf{G}}_{k,m}^H \mathbf{\bar{y}}_m(n) \\ \\ & =\widetilde{\mathbf{G}}_{k,m} \mathbf{\bar{y}}_m(n), 
\end{array}\end{equation}
where we have defined $\widetilde{\mathbf{G}}_{k,m}$ as the following $P_k \times N_{\rm AP}$-dimensional matrix: 
\begin{equation}
\widetilde{\mathbf{G}}_{k,m}= \left(\mathbf{L}_k^H \widehat{\mathbf{G}}_{k,m}^H \widehat{\mathbf{G}}_{k,m}
\mathbf{L}_k\right)^{-1}   \mathbf{L}_k^H \widehat{\mathbf{G}}_{k,m}^H \, .
\end{equation}
The vectors $\widetilde{\mathbf{y}}_{m,k}(n)$, for all $k=1, \ldots, K$, are then sent to the CPU via the backhaul link;
the CPU, finally, forms the following soft estimates of the data vectors transmitted by the users:
\begin{equation}
\widehat{\mathbf{x}}^{\rm UL,cf}_k(n)= \ds \sum_{m=1}^M \widetilde{\mathbf{y}}_{m,k}(n) \; , \quad k=1, \ldots, K \, .
\label{Est_UL_cf1}
\end{equation}
Note that only the soft estimates $ \widetilde{\mathbf{y}}_{m,k}(n)$ are to be transmitted from the APs to the CPU, while channel estimates transmission is not required. 
\subsubsection{UC massive MIMO architecture}
In this case, the signal transmitted by the $k$-th MS is decoded only by the APs in the set ${\cal M}(k)$ . Otherwise stated, the $m$-th AP computes the statistics $\widetilde{\mathbf{y}}_{m,k}(n)$
only for the MSs in ${\cal K}(m)$.
Accordingly, the CPU is able to perform the following soft estimates for the data sent by the $K$ MSs in the system:
\begin{equation}
\widehat{\mathbf{x}}^{\rm UL,uc}_k(n)= \ds \sum_{m \in \mathcal{M}(k)} \widetilde{\mathbf{y}}_{m,k}(n) \; , \quad k=1, \ldots, K \, .
\label{Est_UL_uc1}
\end{equation}
Notice that in this case the backhaul overhed is reduced with respect to the CF case since each AP has to send only the soft estimates of the data received by its associated MSs.

\section{Performance measures and downlink power control} \label{Downlink_power_control_section}
We can note that the CF approach can be obtained as a special case of the UC one by letting $N=K$, i.e. each AP serves all the $K$ users in the system, so ${\cal M}(k)=\left\lbrace 1, \ldots , M\right\rbrace, \; \forall k=1,\ldots , K$. Following this approach, in the downlink power control here explained, we denote as $\eta_{k,m}^{\rm DL}$ the generic scalar coefficient ruling the power transmitted in downlink by the $m$-th AP for the $k$-th MS. From \eqref{eq:received_data_MS_UC}, we have that the achievable rate in downlink for the user $k$ is written as
\begin{equation}
\mathcal{R}_k^{\rm DL}\left(\bbeta\right)=W \log_2 \left|\mathbf{I} + \mathbf{R}_k^{-1} \mathbf{A}_{k,k}\mathbf{A}_{k,k}^H \right| \; ,
\label{eq:ASE_expression}
\end{equation}
where 
\begin{equation}
\mathbf{A}_{k,k}=\sum_{m\in{\cal M}(k)} \mathbf{L}_k^H \sqrt{\eta_{k,m}^{\rm DL}} \mathbf{G}_{k,m}^H  \widehat{\mathbf{G}}_{k,m} \left(\widehat{\mathbf{G}}_{k,m}^H \widehat{\mathbf{G}}_{k,m} \right)^{-1} \mathbf{L}_k \; ,
\label{eq:Akk_DL}
\end{equation}
\begin{equation}
\mathbf{R}_k=\sigma^2_z \mathbf{L}_k^H \mathbf{L}_k + \ds \sum_{j=1, j \neq k}^K \mathbf{A}_{k,j}\mathbf{A}_{k,j}^H \; ,
\label{eq:Rk_DL}
\end{equation}
\begin{equation}
\mathbf{A}_{k,j}=  \mathbf{L}_k^H  \sum_{m\in{\cal M}(j)} \ds \sqrt{\eta_{j,m}^{\rm DL}}\mathbf{G}_{k,m}^H   
\widehat{\mathbf{G}}_{j,m} \left(\widehat{\mathbf{G}}_{j,m}^H \widehat{\mathbf{G}}_{j,m} \right)^{-1} \mathbf{L}_j\; 
\label{eq:Akj_DL}
\end{equation}
and $\bbeta$ the $KM\times 1$ vector collecting the transmit powers in downlink of all APs for all MSs.

The rest of this section will be concerned with the optimization of the downlink transmit powers for the maximization of the system sum-rate and minimum users' rate, subject to maximum power constraints.  Mathematically, the sum-rate maximization problem is formulated as the optimization program:
\begin{subequations}\label{Prob:SumRate}
\begin{align}
&\ds\max_{\bbeta}\;\sum_{k=1}^K\mathcal{R}_{k}^{\rm DL}(\bbeta)\label{Prob:aSumRate1}\\
&\;\textrm{s.t.}\; \sum_{k\in{\cal K}_m}\eta_{k,m}^{\rm DL}\leq P_{{\rm max},m}\;,\forall\;m=1,\ldots,M\label{Prob:bSumRate1}\\
&\;\;\;\quad\eta_{k,m}^{\rm DL}\geq 0\;,\forall\;m=1,\ldots,M,\;k=1,\ldots,K,\label{Prob:cSumRate1}
\end{align}
\end{subequations}
whereas the minimum rate maximization problem is 
\begin{subequations}\label{Prob:MinRate}
\begin{align}
&\ds\max_{\bbeta}\;\min_{1\leq k\leq K}\; \mathcal{R}_{k}^{\rm DL}(\bbeta)\label{Prob:aMinRate}\\
&\;\textrm{s.t.}\; \sum_{k\in{\cal K}_m}\eta_{k,m}^{\rm DL}\leq P_{{\rm max},m}\;,\forall\;m=1,\ldots,M\label{Prob:bMinRate}\\
&\;\;\;\quad\eta_{k,m}^{\rm DL}\geq 0\;,\forall\;m=1,\ldots,M,\;k=1,\ldots,K.\label{Prob:cMinRate}
\end{align}
\end{subequations}
 Both problems have non-concave objective functions, which makes their solution challenging. Moreover, even if the problems were concave, the large number of optimization variables, $KM$, would still pose a significant complexity challenge\footnote{Although polynomial, the best known upper-bound for the complexity of generic convex problems scales with the fourth power of the number of variables, while many classes of convex problems admit a cubic complexity \cite{TalNem2001}.}. In order to face these issues, we will resort to the framework of successive lower-bound maximization, recently introduced in\footnote{In \cite{RazaviyaynSIAM} the method is labeled successive upper-bound minimization, since minimization problems are considered.} \cite{RazaviyaynSIAM}, and briefly reviewed next. 

\subsection{Successive lower-bound maximization}\label{Sec:Optimization}
The main idea of the method is to merge the tools of alternating optimization \cite[Section 2.7]{BertsekasNonLinear} and sequential convex programming \cite{SeqCvxProg78}. To elaborate, consider the generic optimization problem 
\begin{align}\label{Prob:GeneralProb}
\ds\max_{\bx\in{\cal X}} f(\bx)\;,
\end{align}
with $f:\mathbb{R}^{n}\to \mathbb{R}$ a differentiable function, and ${\cal X}$ a compact set. As in the alternating optimization method, the successive lower-bound maximization partitions the variable space into $M$ blocks, $\bx=(\bx_1,\ldots,\bx_M)$, which are cyclically optimized one at a time, while keeping the other variable blocks fixed. This effectively decomposes \eqref{Prob:GeneralProb} into $M$ subproblems, with the generic subproblem stated as
\begin{align}\label{Prob:SubProb}
\ds\max_{\bx_{m}} f(\bx_{m},\bx_{-m})\;,
\end{align}
with $\bx_{-m}$ collecting all variable blocks except the $m$-th. It is proved in \cite[Proposition 2.7.1]{BertsekasNonLinear} that iteratively solving \eqref{Prob:SubProb} monotonically improves the value of the objective of \eqref{Prob:GeneralProb}, and converges to a first-order optimal point if the solution of \eqref{Prob:SubProb} is unique for any $m$, and if ${\cal X}={\cal X}_1\times \ldots \times{\cal X}_M$, with $\bx_{m}\in{\cal X}_{m}$ for all $m$. 

Clearly, alternating optimization proves useful when \eqref{Prob:SubProb} can be solved with minor complexity. If this is not the case, the successive lower-bound maximization method proposes to tackle  \eqref{Prob:SubProb} by means of sequential convex programming. This does not guarantee to globally solve \eqref{Prob:SubProb}, but can lead to a computationally feasible algorithm. Moreover, it is guaranteed to preserve the properties of the alternating optimization method \cite{RazaviyaynSIAM}. The idea of sequential optimization is to tackle a difficult maximization problem by solving a sequence of easier maximization problems. To elaborate, let us denote by $g_i(\bx_{m})$ the $i$-th constraint of \eqref{Prob:SubProb},  for $i=1,\ldots,C$. Then, consider a sequence of approximate problems $\{{\cal P}_{\ell}\}_{\ell}$ with objectives $\{f_{\ell}\}_{\ell}$ and constraint functions $\{g_{i,\ell}\}_{i=1}^C$, such that the following three properties are fulfilled, for all $\ell$:
\begin{enumerate}
\item[(\textbf{P1})] $f_{\ell}(\bx_m)\leq f(\bx_m)$, $g_{i,\ell}(\bx_m)\leq g_{i}(\bx_m)$, for all $i$ and $\bx_m$;
\item[(\textbf{P2})] $f_{\ell}(\bx_m^{(\ell-1)})=f(\bx_m^{(\ell-1)})$, $g_{i,\ell}(\bx_m^{(\ell-1)})=g_{i}(\bx_m^{(\ell-1)})$ with $\bx_m^{(\ell-1)}$ the maximizer of $f_{\ell-1}$;
\item[(\textbf{P3})] $\nabla f_{\ell}(\bx_m^{(\ell-1)})=\nabla f(\bx_m^{(\ell-1)})$, $\nabla g_{i,\ell}(\bx_m^{(\ell-1)})=\nabla g_{i}(\bx_m^{(\ell-1)})$.
\end{enumerate}
In \cite{SeqCvxProg78} (see also \cite{Beck2010,RazaviyaynSIAM}) it is shown that, subject to constraint qualifications, the sequence $\{f(\bx_m^{(\ell)})\}_{\ell}$ of the solutions of the $\ell$-th Problem ${\cal P}_{\ell}$, is monotonically increasing and converges. Moreover, every convergent sequence $\{\bx_m^{(\ell)}\}_{\ell}$ attains a first-order optimal point of the original Problem \eqref{Prob:SubProb}. Thus, the sequential approach enjoys strong optimality properties, fulfilling at the same time the monotonic improvement property for the objective function, and the \gls{kkt} first-order optimality conditions for the original problem. Nevertheless, its applicability hinges on determining suitable lower bounds for the original objective to maximize, which fulfill all three properties \textbf{P1}, \textbf{P2}, \textbf{P3}, while at the same time leading to manageable optimization problems. 

In conclusion, the successive lower-bound maximization method can be formulated as variation of the alternating optimization method, in which each subproblem \eqref{Prob:SubProb} is not globally solved, but instead is tackled by sequential optimization theory. It is proved in \cite{RazaviyaynSIAM} that successive lower-bound maximization has the same optimality properties as the true alternating optimization method, under similar assumptions, even though each subproblem might not be globally solved\footnote{Of course, this holds provided the additional  assumption of the sequential method are fulfilled in each iteration}.
 
\subsection{Sum-rate maximization}\label{Sec:SumRate}
Consider Problem \eqref{Prob:SumRate} and define the variable blocks $\bbeta_{m}$, $m=1,\ldots,M$,  collecting the transmit powers of access point $m$. Then, the sum-rate maximization with respect to the variable block $\bbeta_{m}$ is cast as
\begin{subequations}\label{Prob:SumRateSub}
\begin{align}
&\ds\max_{\bbeta_m}\;\sum_{k=1}^K\mathcal{R}_{k}^{\rm DL}(\bbeta_m,\bbeta_{-m})\label{Prob:aSumRate}\\
&\;\textrm{s.t.}\; \sum_{k\in{\cal K}_m}\eta_{k,m}^{\rm DL}\leq P_{{\rm max},m}\label{Prob:bSumRateSub}\\
&\;\;\;\quad\eta_{k,m}^{\rm DL}\geq 0\;,\forall\;k\in {\cal K}_m\label{Prob:cSumRateSub}\; .
\end{align}
\end{subequations}
The complexity of \eqref{Prob:SumRateSub} is significantly lower than that of \eqref{Prob:SumRate}, since only the $M$ transmit powers of access point $m$ are being optimized. Nevertheless, Problem \eqref{Prob:SumRateSub} is still non-convex, which makes its solution difficult. Indeed, defining 
\begin{equation}
\mathbf{A}_{k,j,m}= \mathbf{L}_k^H  \mathbf{G}_{k,m}^H \widehat{\mathbf{G}}_{j,m} \left(\widehat{\mathbf{G}}_{j,m}^H \widehat{\mathbf{G}}_{j,m} \right)^{-1} \mathbf{L}_j\;,
\label{eq:Akjm_DL}
\end{equation}
the $k$-th user's achievable rate can be expressed as \eqref{Eq:kRate} at the top of next page, 
\begin{figure*}
\begin{align}\label{Eq:kRate}
\mathcal{R}_k^{\rm DL}(\bbeta)=\underbrace{W\log_2\left|\sigma_z^2\bL_k^H\bL_k\!+\!\!\!\sum_{j=1}^K\sum_{m}\sum_{\ell}\sqrt{\eta_{j,m}^{\rm DL}\eta_{j,\ell}^{\rm DL} }\bA_{k,j,m}\bA_{k,j,\ell}^H\right|}_{g_1(\bbeta)}\\
-\underbrace{W\log_2\left|\sigma_z^2\bL_k^H\bL_k\!+\!\!\!\sum_{j=1, j \neq k}^K\!\!\sum_{m}\sum_{\ell}\sqrt{\eta_{j,m}^{\rm DL}\eta_{j,\ell}^{\rm DL} }\bA_{k,j,m}\bA_{k,j,\ell}^H\right|}_{g_2(\bbeta)}\;.\notag
\end{align}
\end{figure*}
which can be seen to be non-concave, also with respect to only the variable block $\bbeta_m$. Thus, following the successive lower-bound maximization, \eqref{Prob:SumRateSub} will be tackled by sequential optimization. To this end, it is necessary to derive a lower-bound of the objective of \eqref{Prob:SumRateSub}, which fulfills Properties \textbf{P1}, \textbf{P2}, and \textbf{P3}, while at the same time leading to a simple optimization problem. To this end, the following lemma proves useful. 
\begin{lemma}\label{Lemma:Conc1}
The function $f:(x,y)\in\mathbb{R}^{2}\to\sqrt{xy}$ is jointly concave in $x$ and $y$, for $x,y>0$.
\end{lemma}
\begin{IEEEproof}
The proof follows upon computing the Hessian of $\sqrt{xy}$ and showing that it is negative semi-definite. Details are omitted for the sake of brevity.
\end{IEEEproof}
Lemma \ref{Lemma:Conc1}, coupled with the facts that the function $\log_2|(\cdot)|$ is matrix-increasing, and that summation preserves concavity, implies that the rate function in \eqref{Eq:kRate} is the difference of two concave functions. This observation is instrumental for the derivation of the desired lower-bound. Indeed, recalling that any concave function is upper-bounded by its Taylor expansion around any given point $\bbeta_{m,0}$, a concave lower-bound of ${\cal R}_k$ is obtained as
\begin{align}\label{Eq:BoundRate}
{\cal R}_k^{\rm DL}(\bbeta)&=g_1(\bbeta_m)-g_2(\bbeta_m)\\
&\geq g_1(\bbeta_m)-g_2(\bbeta_{m,0})-\nabla_{\bbeta_m}^Tg_2\rvert_{\bbeta_{m,0}}(\bbeta-\bbeta_{m,0})\notag\\
&=\widetilde{{\cal R}}_k^{\rm DL}(\bbeta_m,\bbeta_{m,0})\;.\notag
\end{align}
Moreover, it is easy to check that $\widetilde{{\cal R}}_k$ fulfills by construction also properties \textbf{P2} and \textbf{P3} with respect to ${\cal R}_k$. Thus, Problem \eqref{Prob:SumRateSub} can be tackled by the sequential optimization framework, by defining the $\ell$-th problem of the sequence, ${\cal P}_{\ell}$, as the convex optimization program:
\begin{subequations}\label{Prob:SumRateApp}
\begin{align}
&\ds\max_{\bbeta_m}\;\sum_{k=1}^K\widetilde{\mathcal{R}}_{k}(\bbeta_m,\bbeta_{m,0},\bbeta_{-m})\label{Prob:aSumRateApp}\\
&\;\textrm{s.t.}\; \sum_{k\in{\cal K}_m}\eta_{k,m}\leq P_{{\rm max},m}\label{Prob:bSumRateApp}\\
&\;\;\;\quad\eta_{k,m}\geq 0\;,\forall\;k\in {\cal K}_m\label{Prob:cSumRateApp}
\end{align}
\end{subequations}
For any $\bbeta_{m,0}$, Problem \eqref{Prob:MinRateApp} can be  solved by means of standard convex optimization theory, since the objective is concave, and the constraints are affine. The resulting power control procedure can be stated as in Algorithm \ref{Alg:SRMax}. Moreover, based on the general theory reviewed in Section \ref{Sec:Optimization}, the following result holds. 
\begin{algorithm}[t]
\caption{Sum rate maximization}
\begin{algorithmic}[1]
\label{Alg:SRMax}
\STATE Set $i=0$ and choose any feasible ${\bbeta}_{2},\ldots,\bbeta_{M}$;
\REPEAT
\FOR{$m=1\to M$}
\REPEAT
\STATE Choose any feasible ${\bbeta}_{m,0}$;
\STATE Let $\bbeta_{m}^{*}$ be the solution of \eqref{Prob:SumRateApp};
\STATE $\bbeta_{m,0}=\bbeta_{m}^{*}$;
\UNTIL{convergence}
\STATE $\bbeta_{m}=\bbeta_{m}^{*}$;
\ENDFOR
\UNTIL{convergence}
\end{algorithmic}
\end{algorithm} 

\begin{proposition}\label{Prop:SRmax}
After each iteration in Line 6 of Algorithm \ref{Alg:SRMax}, the sum-rate value $\sum_{k=1}^K{\cal R}_k$ is not decreased, and the resulting sequence $\{\sum_{k=1}^K{\cal R}_k\}$ converges. Moreover, every limit point of the sequence $\{\bbeta_m\}_m$ fulfills the \gls{kkt} first-order optimality conditions of Problem \eqref{Prob:SumRateSub}. 
\end{proposition}
Two remarks are now in order. First of all an extreme case of Algorithm \ref{Alg:SRMax} is that in which only one variable block is used, namely optimizing all of the transmit powers simultaneously. In this scenario, Algorithm \ref{Alg:SRMax} reduces to a pure instance of sequential optimization, and no alternating optimization is required. Nevertheless, as already mentioned, the complexity of this approach seems prohibitive for large $M$ and $K$.
Then, another extreme case  is that in which the $KM$ transmit powers $\eta_{k,m}$ are optimized one at a time, thus leading to considering $KM$ variable blocks. The advantage of this approach is that each subproblem \eqref{Prob:SumRateApp} would have only one optimization variable, and thus could be solved in semi-closed form. This brings drastic computational complexity savings and proves to be useful especially in the CF scenario, since in this case each variable block $\bbeta_m$ always has dimension $K$. 
 
\subsection{Minimum rate maximization}
Consider Problem \eqref{Prob:MinRate}. Following similar steps as in Section \ref{Sec:SumRate}, Problem \eqref{Prob:MinRate} with respect to variable block $\bbeta_m$ becomes
\begin{subequations}\label{Prob:SubMinRate}
\begin{align}
&\ds\max_{\bbeta_m}\;\min_{1\leq k\leq K}\; \mathcal{R}_{k}^{\rm DL}(\bbeta_m,\bbeta_{-m})\label{Prob:aSubMinRate}\\
&\;\textrm{s.t.}\; \sum_{k\in{\cal K}_m}\eta_{k,m}^{\rm DL}\leq P_{{\rm max},m}\label{Prob:bSubMinRate}\\
&\;\;\;\quad\eta_{k,m}^{\rm DL}\geq 0\;,\forall k\in {\cal K}_m\label{Prob:cSubMinRate}
\end{align}
\end{subequations}
Besides the difficulties already encountered in the sum-rate scenario, Problem \eqref{Prob:SubMinRate} poses the additional challenge of having a non-differentiable objective due to the $\min(\cdot)$ operator. To circumvent this issue, \eqref{Prob:SubMinRate} can be equivalently reformulated as the program:
\begin{subequations}\label{Prob:MinRate2}
\begin{align}
&\ds\max_{\bbeta_m,t}\;t\label{Prob:aMinRate2}\\
&\;\textrm{s.t.}\; \sum_{k\in{\cal K}_m}\eta_{k,m}^{\rm DL}\leq P_{{\rm max},m}\label{Prob:bMinRate2}\\
&\;\;\;\quad\eta_{k,m}^{\rm DL}\geq 0\;,\forall\;k\in {\cal K}_m\label{Prob:cMinRate2}\\
&\;\quad\;\;\mathcal{R}_{k}^{\rm DL}(\bbeta_m,\bbeta_{-m})\geq t\;,\forall\;k=1,\ldots,K\label{Prob:dMinRate2}\;.
\end{align}
\end{subequations}
At this point, it is possible to tackle \eqref{Prob:MinRate2} by the sequential method. Leveraging again the bound in \eqref{Eq:BoundRate} leads to considering the approximate problem 
\begin{subequations}\label{Prob:MinRateApp}
\begin{align}
&\ds\max_{\bbeta_m,t}\;t\label{Prob:aMinRateApp}\\
&\;\textrm{s.t.}\; \sum_{k\in{\cal K}_m}\eta_{k,m}^{\rm DL}\leq P_{{\rm max},m}\label{Prob:bMinRateApp}\\
&\;\;\;\quad\eta_{k,m}^{\rm DL}\geq 0\;,\forall\;k\in {\cal K}_m\label{Prob:cMinRateApp}\\
&\;\quad\;\;\widetilde{{\cal R}}_k^{\rm DL}(\bbeta_m,\bbeta_{m,0},\bbeta_{-m})\geq t\;,\forall\;k=1,\ldots,K\label{Prob:dMinRateApp}\;.
\end{align}
\end{subequations}
For any $\bbeta_{m,0}$, Problem \eqref{Prob:MinRateApp} can be  solved by means of standard convex optimization theory, since the objective is linear, and the constraints are all convex. The resulting power control procedure can be stated as in Algorithm \ref{Alg:MRMax}, which enjoys similar properties as Algorithm \ref{Alg:SRMax}.

\begin{algorithm}[t]
\caption{Minimum rate maximization}
\begin{algorithmic}[1]
\label{Alg:MRMax}
\STATE Set $i=0$ and choose any feasible ${\bbeta}_{2},\ldots,\bbeta_{M}$;
\REPEAT
\FOR{$m=1\to M$}
\REPEAT
\STATE Choose any feasible ${\bbeta}_{m,0}$;
\STATE Let $\bbeta_{m}^{*}$ be the solution of \eqref{Prob:MinRateApp};
\STATE $\bbeta_{m,0}=\bbeta_{m}^{*}$;
\UNTIL{convergence}
\STATE $\bbeta_{m}=\bbeta_{m}^{*}$;
\ENDFOR
\UNTIL{convergence}
\end{algorithmic}
\end{algorithm}

\section{Uplink power control} \label{Uplink_power_control_section}
Considering again the CF approach as a special case of the UC one, from \eqref{Est_UL_uc1}, for the uplink, upon defining 
\begin{equation}
\mathbf{B}_{k,j}=\ds \sum_{m \in {\cal M}(k)}\widetilde{\mathbf{G}}_{k,m}\mathbf{G}_{j,m}\mathbf{L}_j
\label{eq:Bkj}
\end{equation}
we have that the rate for the $k$-th user is
\begin{equation}\label{Eq:RateUplink}
\mathcal{R}_k^{\rm UL}\left(\widetilde{\bbeta}\right)= \log_2  \left|\mathbf{I}_{P_k}+\eta_k^{\rm UL} \mathbf{\widetilde{R}}_{k}^{-1} \mathbf{B}_{k,k}\mathbf{B}_{k,k}^H \right|,
\end{equation}
where 
\begin{equation}
{\widetilde{\mathbf{R}}}_{k}= \ds \sum_{j=1, j \neq k}^K \eta_j^{\rm UL}\ds \mathbf{B}_{k,j}\mathbf{B}_{k,j}^H+\sigma^2_w \ds \sum_{m\in{\cal M}(k)} \ds \widetilde{\mathbf{G}}_{k,m}\widetilde{\mathbf{G}}_{k,m}^H \; ,
\label{eq:interference_ul}
\end{equation}
and $\widetilde{\bbeta}$ is the $K\times 1$ vector collecting the transmit powers in uplink of all MSs.
Then, the sum-rate maximization problem is stated as 
\begin{subequations}\label{Prob:SumRateUP}
\begin{align}
&\ds\max_{\widetilde{\bbeta}}\;\sum_{k=1}^K\mathcal{R}_{k}^{\rm UL}\left(\widetilde{\bbeta}\right)\label{Prob:aSumRateAppUP}\\
&\;\textrm{s.t.}\; 0\leq \eta_{k}^{\rm UL}\leq P_{{\rm max},k}\,\forall\; k=1,\ldots,K\;,\label{Prob:bSumRateAppUP}
\end{align}
\end{subequations}
while the minimum rate maximization problem is stated as 
\begin{subequations}\label{Prob:MinRateUP}
\begin{align}
&\ds\max_{\widetilde{\bbeta}}\;\min_{1,\ldots,K}\mathcal{R}_{k}^{\rm UL}\left(\widetilde{\bbeta}\right)\label{Prob:aMinRateAppUP}\\
&\;\textrm{s.t.}\; 0\leq \eta_{k}^{\rm UL}\leq P_{{\rm max},k}\,\forall\; k=1,\ldots,K\; .\label{Prob:bMinRateAppUP}
\end{align}
\end{subequations}
Based on \eqref{Eq:RateUplink}, it is possible to develop power control algorithms for sum-rate and minimum rate maximization, leveraging the sequential optimization framework, as done in the downlink scenario. Indeed, also in the uplink case, we observe that the $k$-th user's rate \eqref{Eq:RateUplink} can be written as the difference of two concave functions, namely
\begin{align}\label{Eq:RateUplink2}
\mathcal{R}_k^{\rm UL}\left(\widetilde{\bbeta}\right)&=\underbrace{\log_{2}\left|\bG_{k}+\sum_{j=1}^{K}\eta_{j}^{\rm UL} \mathbf{B}_{k,j}\mathbf{B}_{k,j}^H \right|}_{g_{1}\left(\widetilde{\bbeta}\right)}\notag\\
&-\underbrace{\log_{2}\left|\bG_{k}+\sum_{j=1, j\neq k}^{K}\eta_{j}^{\rm UL} \mathbf{B}_{k,j}\mathbf{B}_{k,j}^H \right|}_{g_{2}\left(\widetilde{\bbeta}\right)}\;,
\end{align}
wherein $\bG_{k}=\sigma^2_w \ds \sum_{m\in{\cal M}(k)} \ds \widetilde{\mathbf{G}}_{k,m}\widetilde{\mathbf{G}}_{k,m}^H$ for all $k=1,\ldots,K$. 
Now, it is clear that both $g_{1}(\cdot)$ and $g_{2}(\cdot)$ are concave functions of $\widetilde{\bbeta}$ and thus \eqref{Eq:RateUplink2} shows that the $k$-th user's rate can be once again written as the difference of two concave functions. As a consequence, for all $k=1,\ldots,K$, a lower-bound of the $k$-th user's rate, which fulfills all three properties of the sequential optimization method, say $\widetilde{\mathcal{R}}_k^{\rm UL}\left(\widetilde{\bbeta}\right)$, is given by \eqref{Eq:BoundRate}, in which $g_{1}$ and $g_{2}$ take the expression in \eqref{Eq:RateUplink2} above. 
\begin{remark}\label{Rem:Uplink}
In the downlink case the number of optimization variables was $KM$, with $M>K$, and this made it convenient, for complexity reasons, to partition the variable space into multiple blocks of variables that were alternatively optimized. On the other hand, in the uplink case we only have $K$ variables, and this makes it practically feasible to consider only one variable block, thus optimizing all variables at the same time. In the sequel, the focus will be on this case, but we stress that it, if desired, the optimization algorithms can be straightforwardly extended to the scenario in which multiple optimization blocks are defined and iteratively optimized.  
\end{remark}
Keeping Remark \ref{Rem:Uplink} in mind, both sum-rate maximization and minimum rate maximization can be performed by similar algorithms as Algorithms \ref{Alg:SRMax} and \ref{Alg:MRMax}, respectively, in which the auxiliary problem to be solved within each iteration are stated as
\begin{subequations}\label{Prob:SumRateAppUP}
\begin{align}
&\ds\max_{\widetilde{\bbeta}}\;\sum_{k=1}^K\widetilde{\mathcal{R}}_{k}^{\rm UL}\left(\widetilde{\bbeta},\widetilde{\bbeta}_{0} \right)\label{Prob:aSumRateAppUP2}\\
&\;\textrm{s.t.}\; 0\leq \eta_{k}^{\rm UL}\leq P_{{\rm max},k}\,\forall\; k=1,\ldots,K\label{Prob:bSumRateAppUP2}
\end{align}
\end{subequations}
for sum-rate maximization, and as 
\begin{subequations}\label{Prob:MinRateAppUP}
\begin{align}
&\ds\max_{\widetilde{\bbeta}_m,t}\;t\label{Prob:aMinRateAppUP2}\\
&\;\textrm{s.t.}\; 0\leq\eta_{k}^{\rm UL}\leq P_{{\rm max},k}\;\forall\;k=1,\ldots,K\label{Prob:bMinRateAppUP2}\\
&\;\quad\;\;\widetilde{{\cal R}}_{k}^{\rm UL}\left(\widetilde{\bbeta},\widetilde{\bbeta}_{0} \right)\geq t\;,\forall\;k=1,\ldots,K\label{Prob:cMinRateAppUP}\;,
\end{align}
\end{subequations}
for minimum rate maximization. Similar optimality properties as in the downlink case hold. 

\begin{figure}[t]
\centering
\includegraphics[scale=0.5]{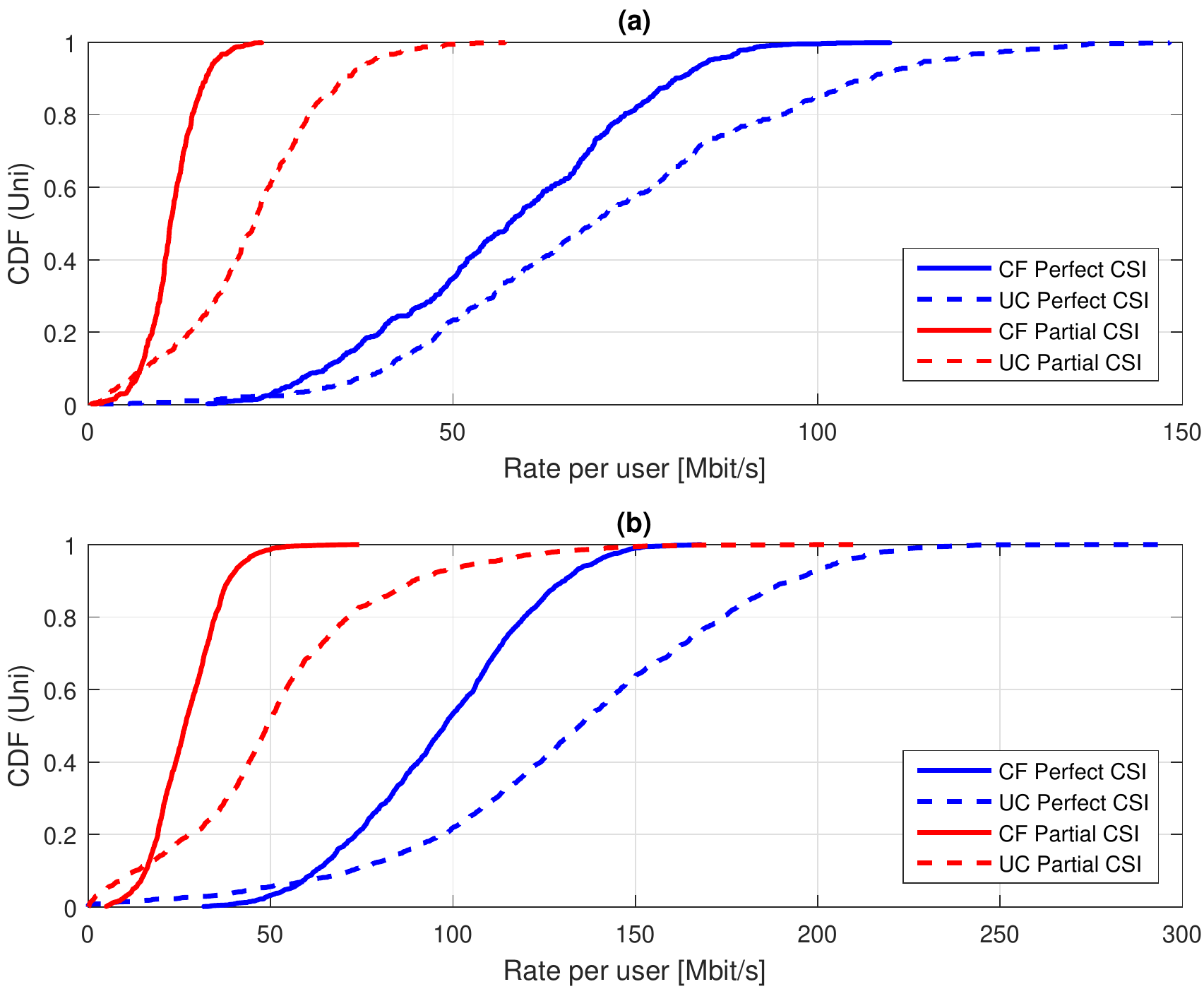}
\caption{CDF of rate per user in downlink with uniform power allocation for a high density scenario in subfigure (a) and for a low density scenario in subfigure (b). Parameters: (a) $M=80$, $K=15$, $N=6$, and $\tau_p=16$; (b) $M=50$, $K=5$, $N=2$, and $\tau_p=8$.}
\label{Fig:CDF_Uni_DL}
\end{figure}

\begin{figure}[t]
\centering
\includegraphics[scale=0.5]{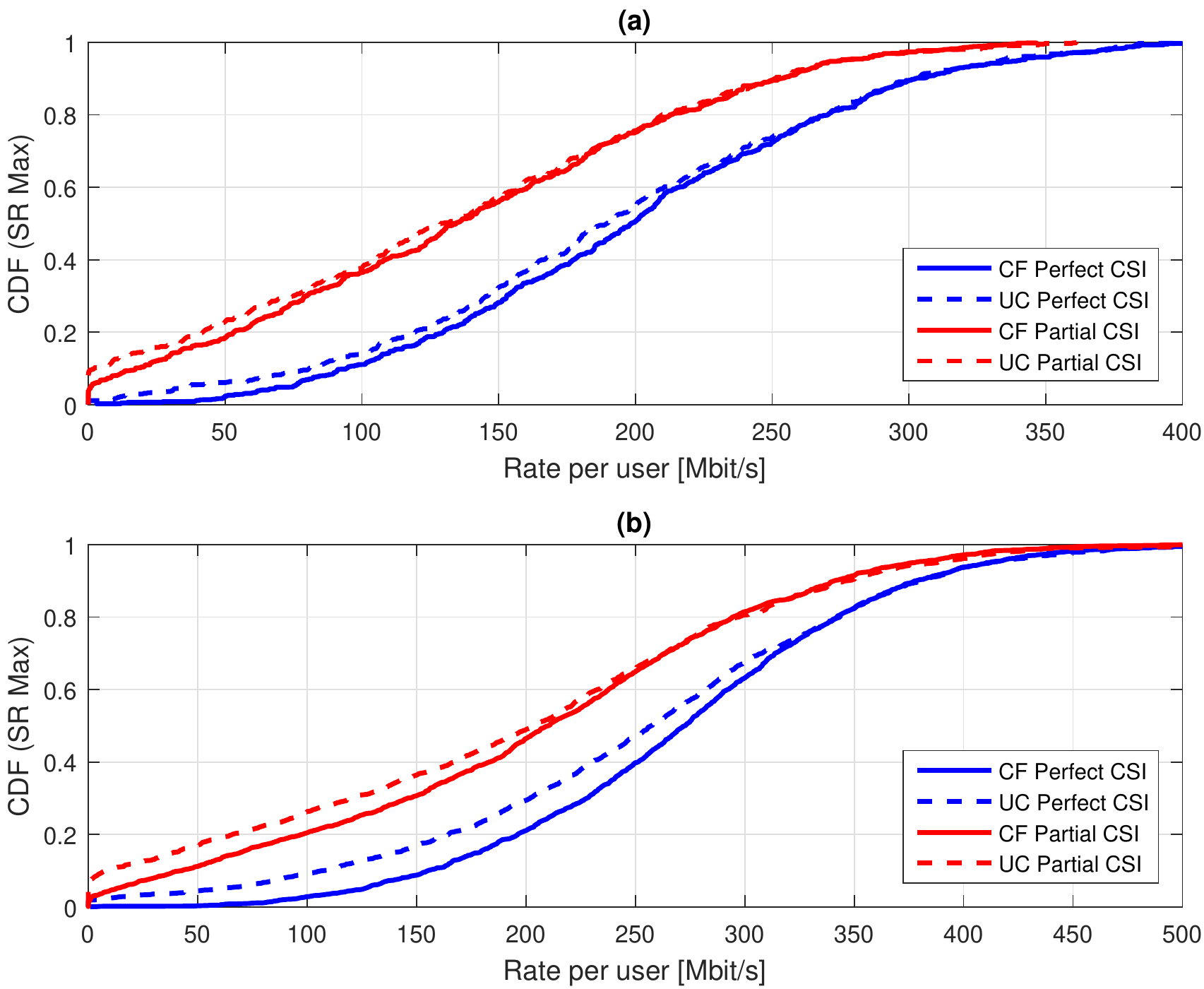}
\caption{CDF of rate per user in downlink with sum-rate maximizing power allocation for a high density scenario in subfigure (a) and for a low density scenario in subfigure (b). Parameters: (a) $M=80$, $K=15$, $N=6$, and $\tau_p=16$; (b) $M=50$, $K=5$, $N=2$, and $\tau_p=8$.}
\label{Fig:CDF_SRMax_DL}
\end{figure}

\begin{figure}[t]
\centering
\includegraphics[scale=0.5]{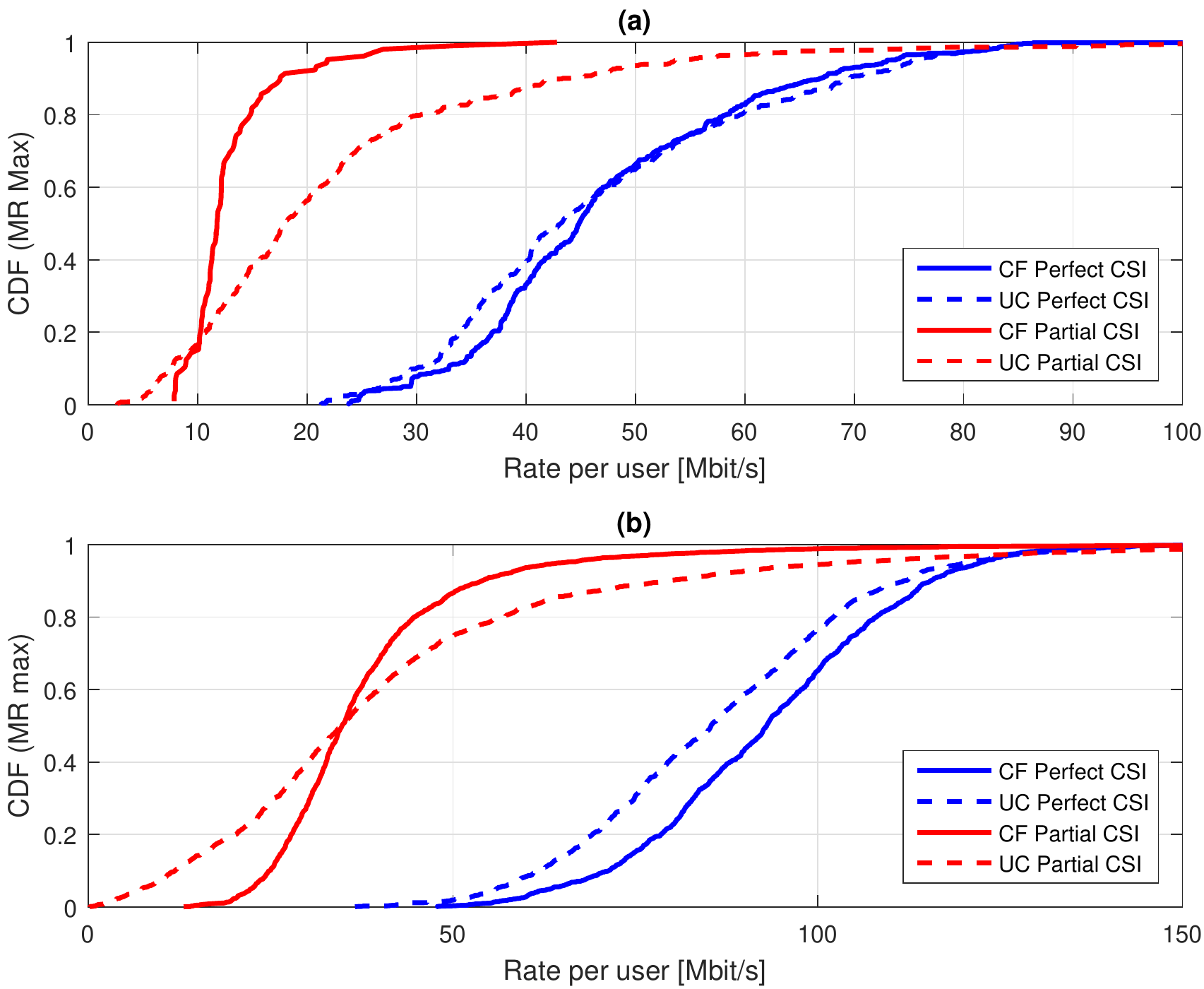}
\caption{CDF of rate per user in downlink with minimum-rate maximizing power allocation for a high density scenario in subfigure (a) and for a low density scenario in subfigure (b). Parameters: (a) $M=80$, $K=15$, $N=6$, and $\tau_p=16$; (b) $M=50$, $K=5$, $N=2$, and $\tau_p=8$.}
\label{Fig:CDF_MRMax_DL}
\end{figure}

\begin{figure}[t]
\centering
\includegraphics[scale=0.5]{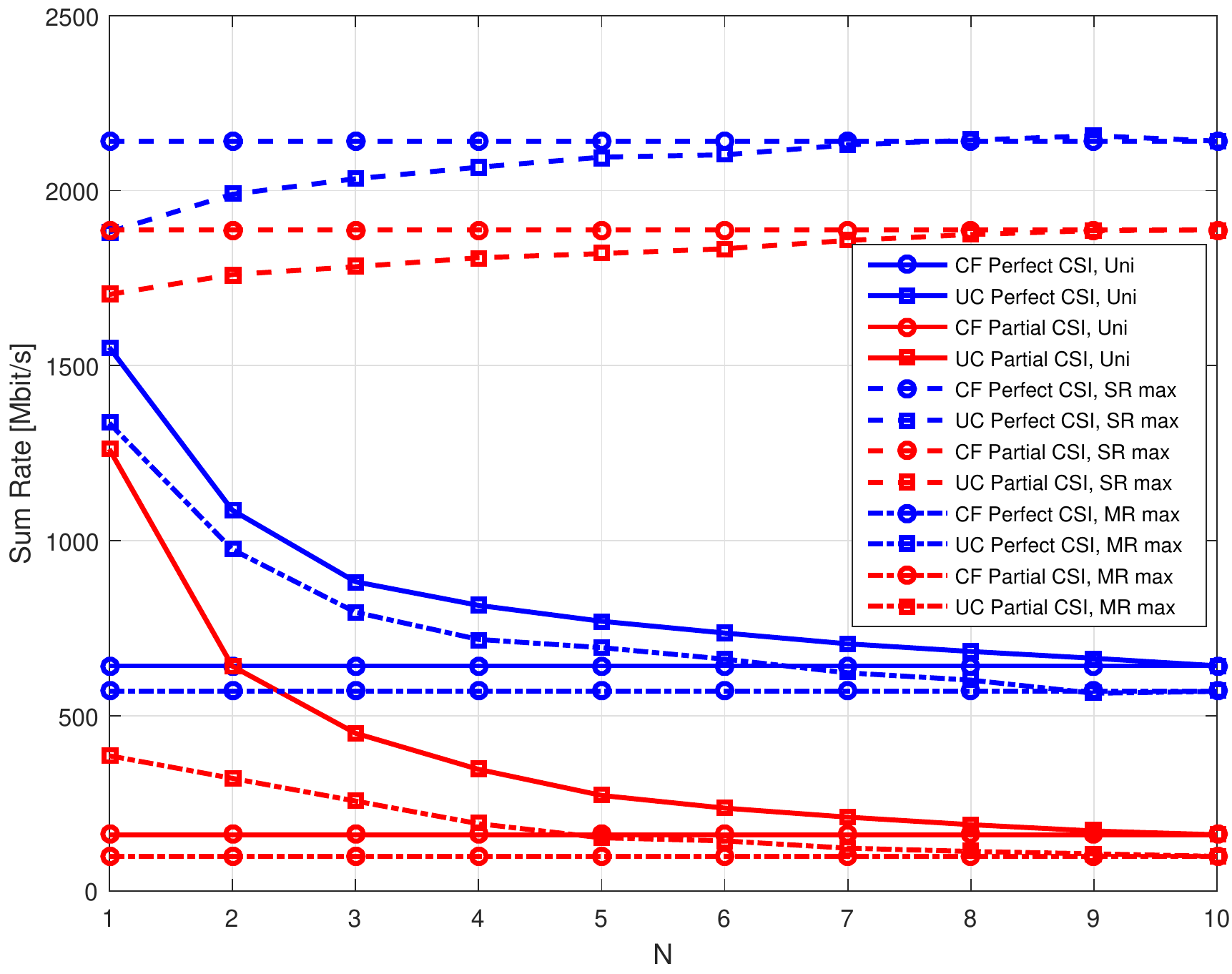}
\caption{Sum rate of the system in downlink versus $N$. Parameters: $M=60$, $K=10$, and $\tau_p=16$; }
\label{Fig:Sum_Rate_N_DL}
\end{figure}

\section{Numerical Results} \label{Numerical_section}
In our simulation setup, we consider a communication bandwidth of $W = 20$ MHz centered
over the carrier frequency $f_0=1.9$ GHz. The antenna height at the AP is $15$ m and at the MS is $1.65$ m. The standard deviation of the shadow fading is $\sigma_{\rm sh}=8$ dB, the parameters for the three slope path loss model in \eqref{path_loss} are $d_1=50$ m and $d_0=10$ m, the parameter $\delta$ in \eqref{shadowing} is 0.5 and the correlation distance in \eqref{shadowing_corr} is $d_{\rm decorr}=100$ m. The additive thermal noise is assumed to have a power spectral density of $-174$ dBm/Hz, while the front-end receiver at the AP and at the MS is assumed to have a noise figure of $9$ dB. In order to emulate an infinite area and to avoid boundary effects, the square area is wrapped around as in reference \cite{Ngo_CellFree2017}. The shown results come from an average over 100 random scenario realizations with independent MSs and APs locations and channels. 
We quantitatively study and compare the performance of the CF and UC massive MIMO architectures.  We consider a square area of $1000 \times 1000$ sqm.; we assume $N_{\rm AP}=4$, $N_{\rm MS}=2$ and the multiplexing order per user is $P_k=2, \; \forall k=1,\ldots,K$. 
For the PM channel estimation, we use maximum-length-sequences (pseudo-noise) with length $\tau_p$ and the uplink transmit power for channel estimation is $p_k=100$ mW, $\forall k=1,\ldots,K$.
We will consider as performance measure the achievable rate per user, and the sum rate of the system in uplink and downlink, both measured in bit/s, implementing the power allocation strategies proposed in the paper and comparing them with the uniform power allocation strategy.
For the uniform power allocation in downlink,  we simply assume that each AP uniformly divides its maximum power among the MSs that it serves in the system. So, for the CF massive MIMO architecture we have
\begin{equation}
\eta_{k,m}^{\rm DL,cf}= \ds \frac{P_{{\rm max},m}}{K \mbox{tr}\left(
\mathbf{Q}_{k,m} \mathbf{Q}^{H}_{k,m}
\right)} \; ,
\end{equation}
and for the UC massive MIMO architecture we have
\begin{equation}
\eta_{k,m}^{\rm DL,uc}=\left\lbrace
\begin{array}{lll}
\ds \frac{P_{{\rm max},m}}{\text{card}\left[{\cal K}(m)\right] \mbox{tr}\left(
\mathbf{Q}_{k,m} \mathbf{Q}^{H}_{k,m}
\right)} & & \text{if} \; \; k \in \mathcal{K}(m) \\
0 & & \text{otherwise} \; ,
\end{array} \right.
\end{equation} 
where $\text{card}\left[{\cal K}(m)\right]$ denotes the cardinality of the set ${\cal K}(m)$.
For the uniform power allocation in uplink, we assume that each MSs transmits with its maximum power, so in the CF and UC massive MIMO architecture we assume 
\begin{equation}
\eta_{k}^{\rm UL,cf}=\eta_{k}^{\rm UL,uc}= \ds \frac{P_{{\rm max},k}}{N_{\rm MS}}\; ,\;  \forall k=1,\ldots,K.
\end{equation}

\begin{figure}[t]
\centering
\includegraphics[scale=0.5]{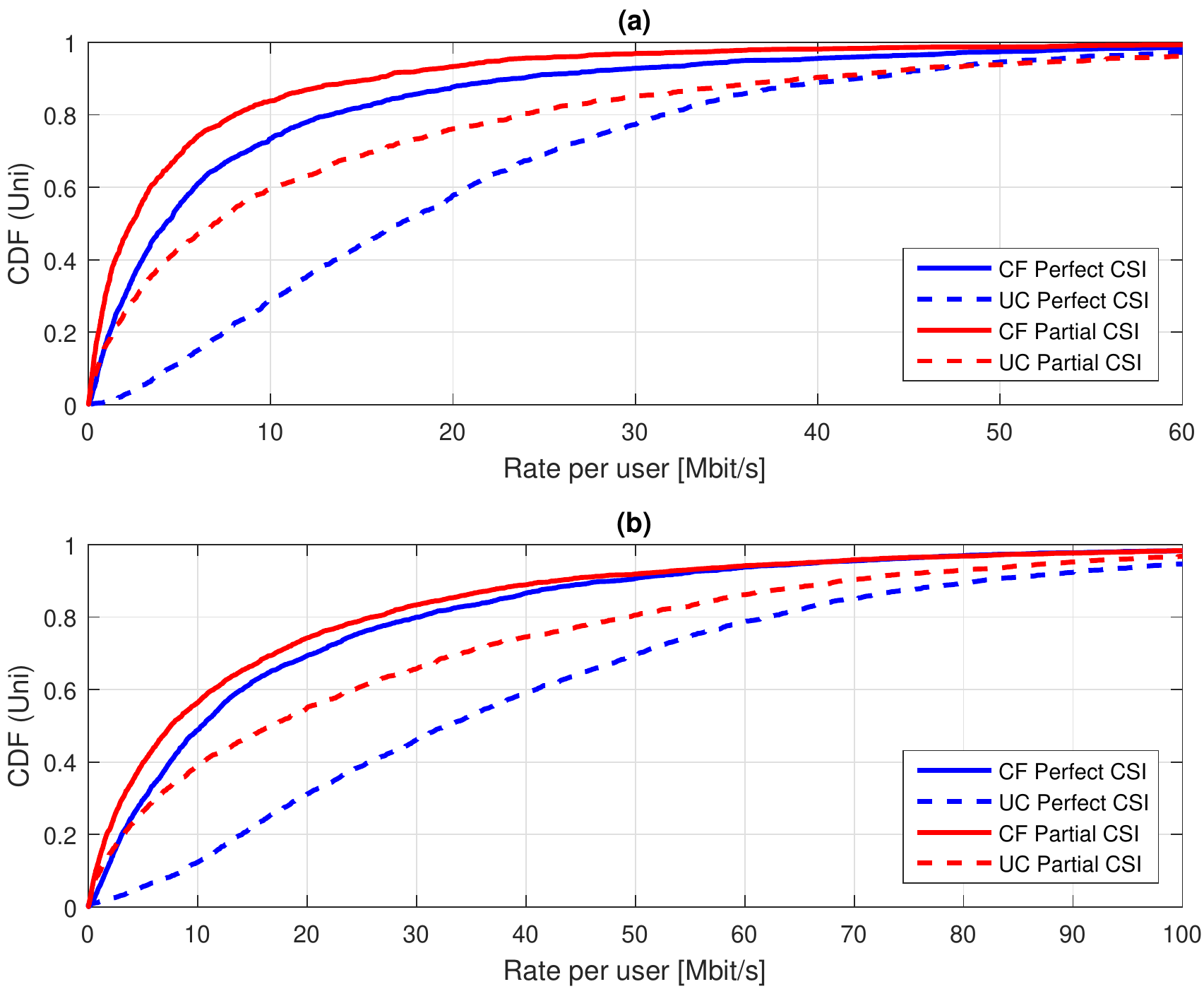}
\caption{CDF of rate per user in uplink with uniform power allocation for a high density scenario in subfigure (a) and for a low density scenario in subfigure (b). Parameters: (a) $M=80$, $K=15$, $N=6$, and $\tau_p=16$; (b) $M=50$, $K=5$, $N=2$, and $\tau_p=8$.}
\label{Fig:CDF_Uni_UL}
\end{figure}

\begin{figure}[t]
\centering
\includegraphics[scale=0.5]{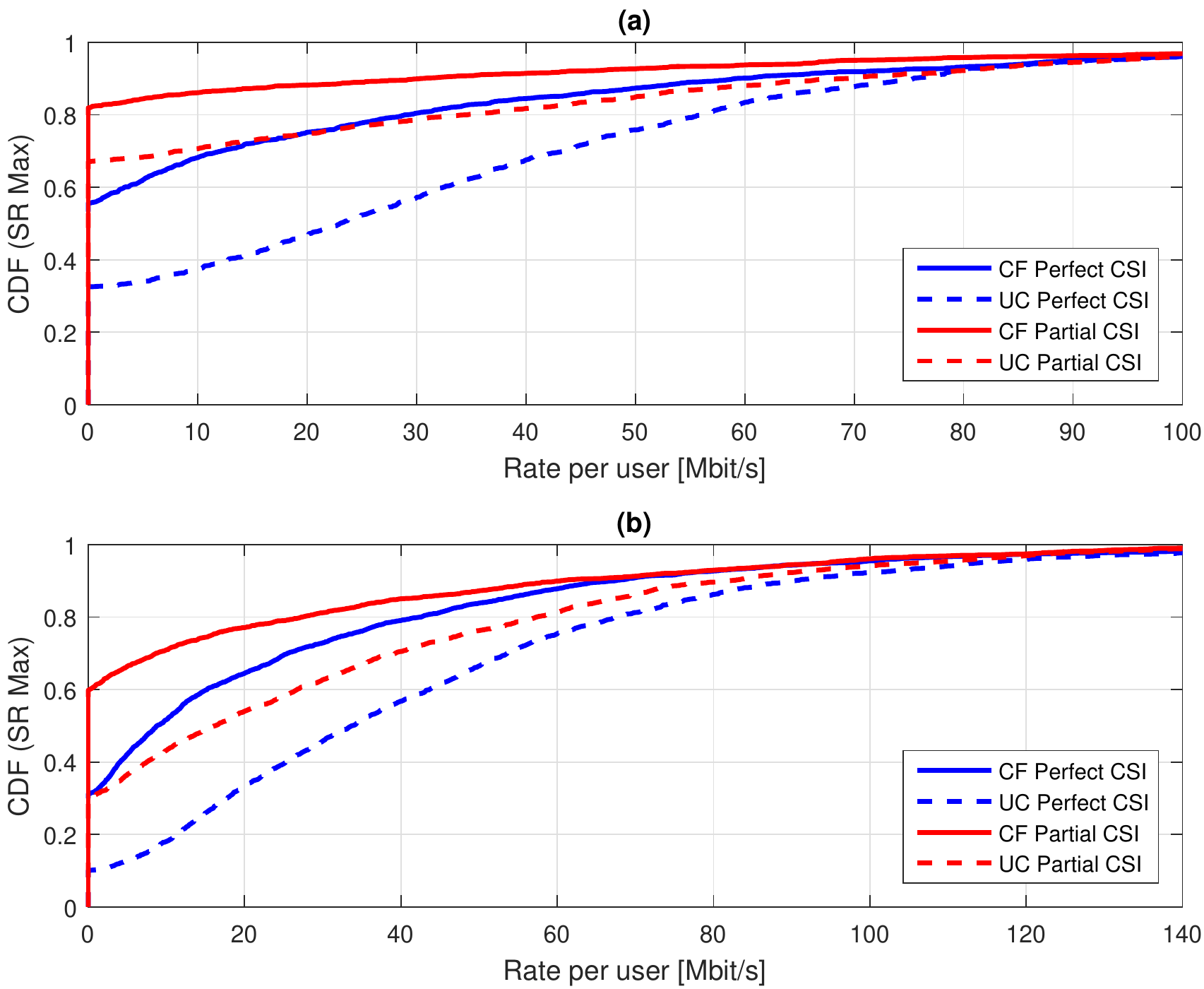}
\caption{CDF of rate per user in uplink with sum-rate maximizing power allocation for a high density scenario in subfigure (a) and for a low density scenario in subfigure (b). Parameters: (a) $M=80$, $K=15$, $N=6$, and $\tau_p=16$; (b) $M=50$, $K=5$, $N=2$, and $\tau_p=8$.}
\label{Fig:CDF_SRMax_UL}
\end{figure}

\begin{figure}[t]
\centering
\includegraphics[scale=0.5]{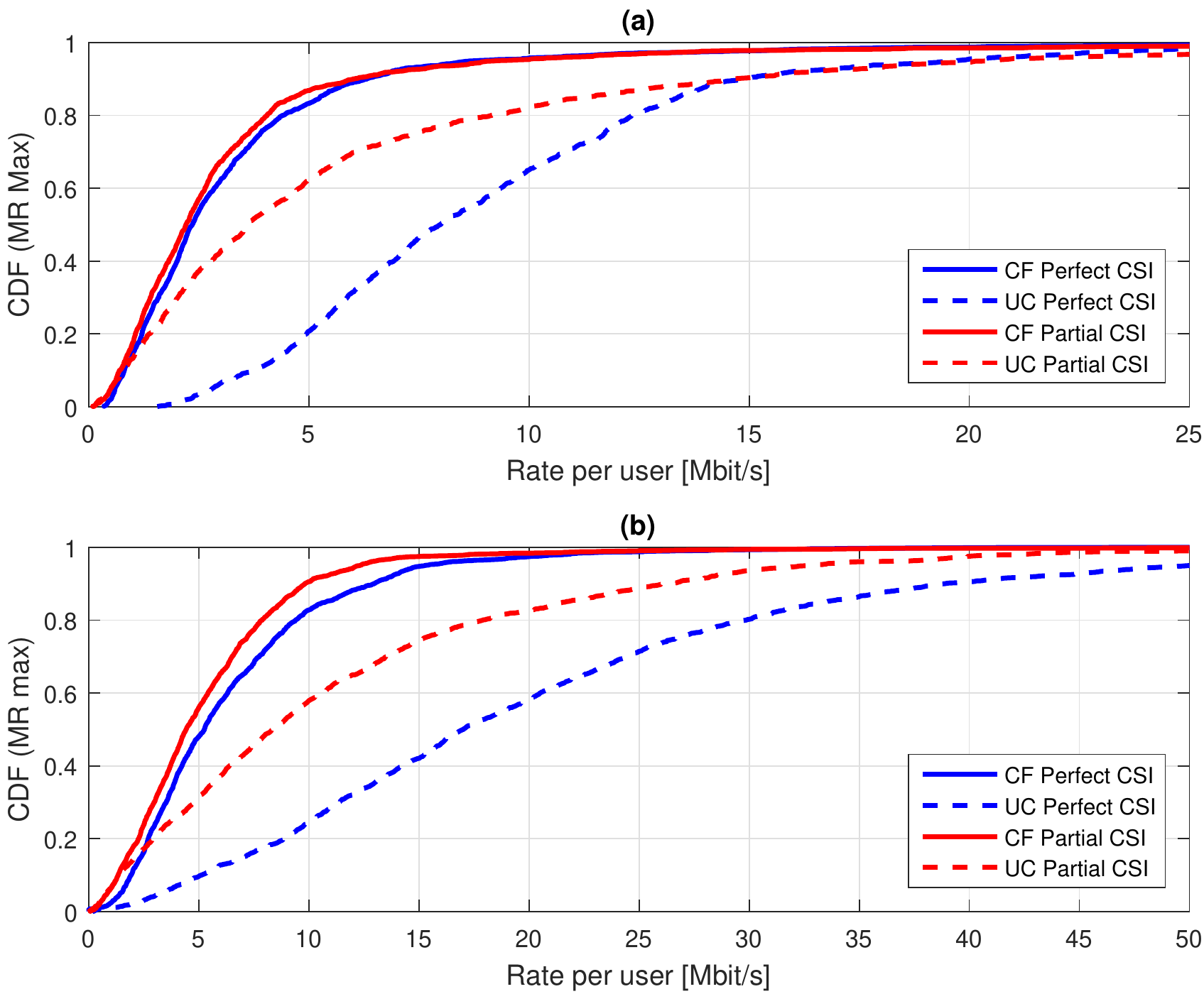}
\caption{CDF of rate per user in uplink with minimum-rate maximizing power allocation for a high density scenario in subfigure (a) and for a low density scenario in subfigure (b). Parameters: (a) $M=80$, $K=15$, $N=6$, and $\tau_p=16$; (b) $M=50$, $K=5$, $N=2$, and $\tau_p=8$.}
\label{Fig:CDF_MRMax_UL}
\end{figure}

\begin{figure}[t]
\centering
\includegraphics[scale=0.5]{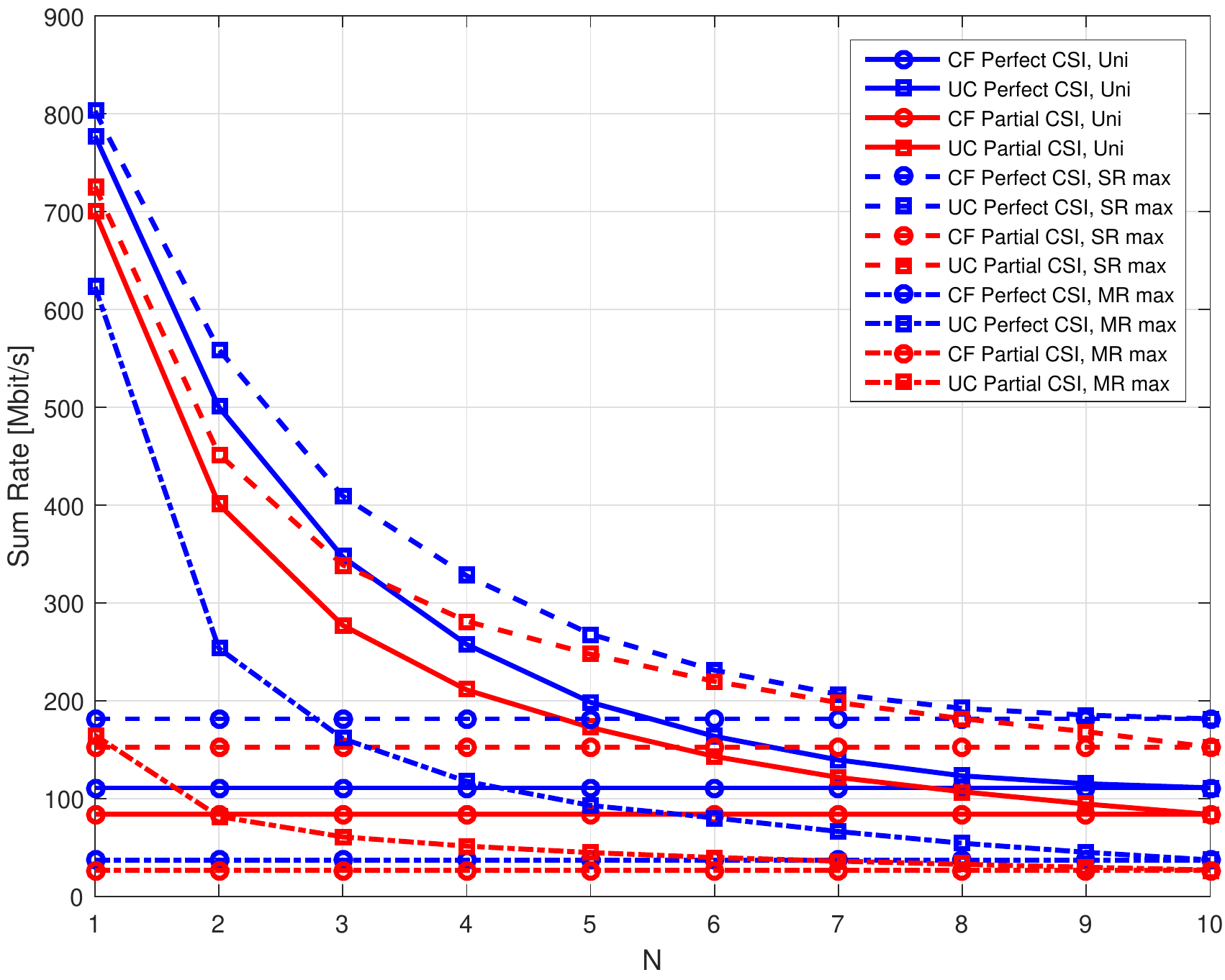}
\caption{Sum rate of the system in uplink versus $N$. Parameters: $M=60$, $K=10$, and $\tau_p=16$; }
\label{Fig:Sum_Rate_N_UL}
\end{figure}

We start considering the performance measures in downlink. Fig. \ref{Fig:CDF_Uni_DL} shows the cumulative distribution functions (CDFs) of the rate per user in downlink for the CF and UC approaches for the case in which uniform power allocation (Uni) is used. Fig. \ref{Fig:CDF_SRMax_DL} reports the CDFs of the rate per user in downlink for the CF and UC approaches for the case in which sum-rate maximizing power allocation (SR Max) is used and Fig. \ref{Fig:CDF_MRMax_DL} reports CDFs of the rate per user in downlink for the CF and UC approaches for the case in which minimum-rate maximizing power allocation (MR Max) is used. Both the cases of perfect channel state information (CSI) and of partial (i.e., estimated) CSI are considered. In Figs. \ref{Fig:CDF_Uni_DL}, \ref{Fig:CDF_SRMax_DL}, and \ref{Fig:CDF_MRMax_DL} we compare the results of the rate per user in the high and low density scenarios, for the high density scenario we assume $M=80$, $K=15$, $N=6$, and $\tau_p=16$, while for the low density scenario we consider $M=50$, $K=5$, $N=2$, and $\tau_p=8$. Fig. \ref{Fig:Sum_Rate_N_DL} shows the sum-rate of the system in uplink versus $N$; in this simulation we assume $M=60$, $K=10$, and $\tau_p=16$. In these figures, we set the maximum power available at each AP at 200 mW, i.e. $P_{{\rm max},m}=200$ mW, $\forall m=1,\ldots, M$. 
Inspecting the figures, we can see that the UC approach outperforms the CF one in the cases of Uni and MR Max, both in the cases of high and low density scenarios. In the case of SR max we can see that the CF approach outperforms the UC one, but the difference between the two approaches is rather limited. This behaviour is also confirmed by Fig. \ref{Fig:Sum_Rate_N_DL}. 

With regard to the uplink, in Fig. \ref{Fig:CDF_Uni_UL} we report the CDFs of the rate per user for the CF and UC approaches for the case in which uniform power allocation is used. Fig. \ref{Fig:CDF_SRMax_UL} shows the CDFs of the rate per user in uplink for the CF and UC approaches for the case in which sum-rate maximizing power allocation is used and Fig. \ref{Fig:CDF_MRMax_UL} reports the CDFs of the rate per user in uplink for the CF and UC approaches for the case in which minimum-rate maximizing power allocation is used. In Figs. \ref{Fig:CDF_Uni_UL}, \ref{Fig:CDF_SRMax_UL}, and \ref{Fig:CDF_MRMax_UL} we compare the results of the rate per user in the high and low density scenarios, the parameters used here are the same as in Figs. \ref{Fig:CDF_Uni_DL}, \ref{Fig:CDF_SRMax_DL}, and \ref{Fig:CDF_MRMax_DL} . Fig. \ref{Fig:Sum_Rate_N_UL} shows the system uplink sum-rate versus $N$, the number of MSs served from each AP, when the number of users is $K=10$. In these figures, we assume that the maximum power available at each MS is 100 mW, i.e. $P_{{\rm max},k}=100$ mW, $\forall k=1,\ldots, K$. The results show that on the uplink, the UC approach outperforms the CF approach for all the power allocation strategies here considered. In particular, there are situations in which the UC approach guarantees many-fold improvements with respect to the CF strategy.

\section{Conclusions} \label{conclusions_section}
The paper has focused on  the recently introduced CF massive MIMO architecture.
First of all, we have extended the CF approach to the case in which both the APs and the MSs are equipped with multiple antennas, and we have proposed the use of a channel-inverting beamforming scheme that does not require channel estimation at the MSs. Then, we have contrasted the CF architecture with the UC approach wherein each AP only decodes a pre-assigned number of MSs.  We have  proposed two power allocation strategies for the uplink and downlink, both for the CF and the UC cases. The first one is a sum-rate maximizing power allocation strategy, aimed at maximizing performance of the system in terms of overall data-rate, while the second one is a minimum-rate maximizing power allocation, aimed at maximizing performance of the system in terms of fairness. We compare the results of the power allocation strategies here proposed with the case of uniform power allocation. Results have shown that the UC approach generally outperforms the CF one, especially on the uplink.  The UC approach thus exhibits in many relevant practical situations better performance than the CF approach, which motivates further investigation from the authors. Relevant research topics worth being investigated are, among others, the following: (a) the consideration of the CF massive MIMO approach at millimeter wave frequencies (preliminary results on this are reported in \cite{Alonzo_Buzzi_PIMRC}); (b)the suitability of a UC architecture to support ultra-reliable low-latency communications; and, finally, (c) the coupling of cell-free massive MIMO architectures with 5G-and-beyond multiple access schemes such as the well-known non-orthogonal multiple access (NOMA). These topics form the object of current research.

\bibliographystyle{IEEEtran}
\bibliography{FracProg_SB,finalRefs,references}

\end{document}